\documentclass[iop,revtex4]{emulateapj}
\newcommand{\bvec}[1]{\textbf{#1}}

\newcommand{\mytilde}{\raise.19ex\hbox{$\scriptstyle\sim$}}
\newcommand{\ciza}{CIZA~J2242.8+5301}
\newcommand{\kms}{$\mbox{km}~\mbox{s}^{-1}$}

\newcommand{\hubblem}{h_{70}^{-1}}

\newcommand{\solarm}{$\hubblem 10^{14}M_{\sun}$}
\newcommand{\elgordo}{ACT-CL~J0102$-$4915}
\usepackage{graphicx}

\shorttitle{WEAK-LENSING STUDY OF CIZA J2242.8+5301}
\shortauthors{Jee et al.}

\begin{document}

\title{$MC^2$: CONSTRAINING THE  DARK MATTER DISTRIBUTION OF THE VIOLENT MERGING GALAXY CLUSTER \ciza~ BY PIERCING THROUGH THE MILKY WAY }

\author{M.~JAMES~JEE\altaffilmark{1},  ANDRA STROE\altaffilmark{2}, WILLIAM DAWSON\altaffilmark{3}, DAVID WITTMAN\altaffilmark{1}, HENK HOEKSTRA\altaffilmark{2},   MARCUS BR{\"U}GGEN\altaffilmark{4}, HUUB R{\"O}TTGGERING\altaffilmark{2}, DAVID SOBRAL\altaffilmark{2,5,6},  REINOUT J. van WEEREN\altaffilmark{7}  }

\altaffiltext{1}{Department of Physics, University of California, Davis, One Shields Avenue, Davis, CA 95616, USA}
\altaffiltext{2}{Leiden Observatory, Leiden University, PO Box 9513, NL-2300 RA Leiden, Netherlands}
\altaffiltext{3}{Lawrence Livermore National Laboratory, P.O. Box 808 L-210, Livermore, CA, 94551, USA }
\altaffiltext{4}{Hamburger Sternwarte, Gojenbergsweg 112, 21029 Hamburg, Germany}
\altaffiltext{5}{Instituto de Astrof\'{\i}sica e Ci\^{e}ncias do Espa\c{c}o, Universidade de Lisboa, OAL, Tapada da Ajuda, PT1349-018 Lisboa, Portugal}
\altaffiltext{6}{Centro de Astronomia e Astrof\'{\i}sica da Universidade de Lisboa, Observat\'{o}rio Astron\'{o}mico de Lisboa, Tapada da Ajuda, 1349-018 Lisboa, Portugal}
\altaffiltext{7}{Harvard Smithsonian Center for Astrophysics, 60 Garden Street Cambridge, MA 02138, USA}

\begin{abstract}
The galaxy cluster \ciza~is a merging system with a prominent ($\mytilde2$~Mpc long) radio relic, which together with the morphology of the X-ray emission provides strong evidence for a violent collision  along the north-south axis. We present our constraints on the dark matter distribution of this unusual system using  Subaru and Canada-France-Hawaii Telescope (CFHT) imaging data. Measuring a high S/N lensing signal from this cluster is potentially a challenging task because of its proximity  to the Milky Way plane ($\left | b \right | \sim5\degr$).
We overcome this challenge with careful observation planning and systematics control, which enables us to successfully map the dark matter distribution of the cluster with high fidelity. The resulting mass map shows that the mass distribution of \ciza ~is highly elongated along the north-south merger axis inferred from the
orientation of the radio relics. Based on our mass reconstruction, we identify two sub-clusters, which coincide with the cluster galaxy distributions. We determine their masses using Markov-Chain-Monte-Carlo analysis by simultaneously fitting two Navarro-Frenk-White (NFW) halos without fixing their centroids. The resulting masses of the northern and southern systems are $M_{200}=11.0_{-3.2}^{+3.7}\times10^{14} M_{\sun}$ and $9.8_{-2.5}^{+3.8}\times10^{14} M_{\sun}$, respectively, indicating that we are witnessing a post-collision of two giant  systems of nearly equal mass. 
When the mass and galaxy centroids are compared in detail,  we detect $\mytilde 1\arcmin$ ($\mytilde190$ kpc) offsets in both northern and southern sub-clusters.
After investigating the statistical significance of the offsets by bootstrapping both mass and galaxy centroids, we find that the galaxy luminosity-mass offset
for the northern clump is statistically significant at the $\gtrsim2\sigma$ level whereas the detection is only marginal for the southern sub-cluster in part because of a relatively large mass centroid error.
We conclude that it is yet premature to uniquely attribute the galaxy-mass misalignment to self-interaction of dark matter and discuss caveats.
\end{abstract}


\section{INTRODUCTION} \label{section_introduction}
Merging galaxy clusters are receiving growing attention because of their potential to provide constraints on properties of dark matter, which is gravitationally the most dominant constituent of galaxy clusters. The capability of weak lensing in constraining the mass distribution of a foreground lens based on its gravitational impact on the shapes of background galaxies provides a unique opportunity to learn from the difference in distribution between dark matter and  baryonic components (galaxies and gas). If collisionless, dark matter particles will distribute nearly in the same way as galaxies do. The famous example supporting this standard paradigm is the ``Bullet Cluster" 1E0657-56 (Clowe et al. 2006), which exhibits no significant difference between the locations of the cluster galaxies and mass peaks. However, there have been reports on some exceptional cases, where weak-lensing reveals mass distributions apparently at odds with the conventional paradigm of collisionless dark matter.
For example, in A520 (Mahdavi et al. 2007; Jee et al. 2012; Jee et al. 2014a) and A2744 (Merten et al. 2011), significant ($\gtrsim 6~\sigma$ in case of A520) mass peaks
without any strong concentration of luminous cluster galaxies are detected. 
In addition, in the case of the Musket Ball cluster (Dawson et al. 2012), a marginal detection of a weak-lensing mass peak trailing galaxies is claimed (Dawson 2013).
Kahlhoefer et al. (2013) have shown that such observational anomalies may be produced by self-interacting dark matter (SIDM). 
Thus they open up interesting research opportunities to study a collection of these rare systems and put meaningful constraints on the properties of dark matter. 

We recently launched a Merging Cluster Collaboration\footnote{http://www.mergingclustercollaboration.org} ($MC^2$; PI. W. Dawson) project with aims to
enhance our understanding of cluster physics and constrain properties of dark matter from systematically analyzing a number of prominent merging clusters.
Combining data from different instruments (e.g., X-ray, spectroscopy, weak-lensing, radio relic observation, etc.) will greatly reduce the parameter space that we need to explore in setting up initial conditions. The simulation of the $MC^2$ collaboration will include various models of dark matter (e.g., Rocha et al. 2013). Iterative comparisons between the simulations and observations will enable us to constrain properties of dark matter with unprecedented precision.

\begin{figure*}
\includegraphics[width=18cm]{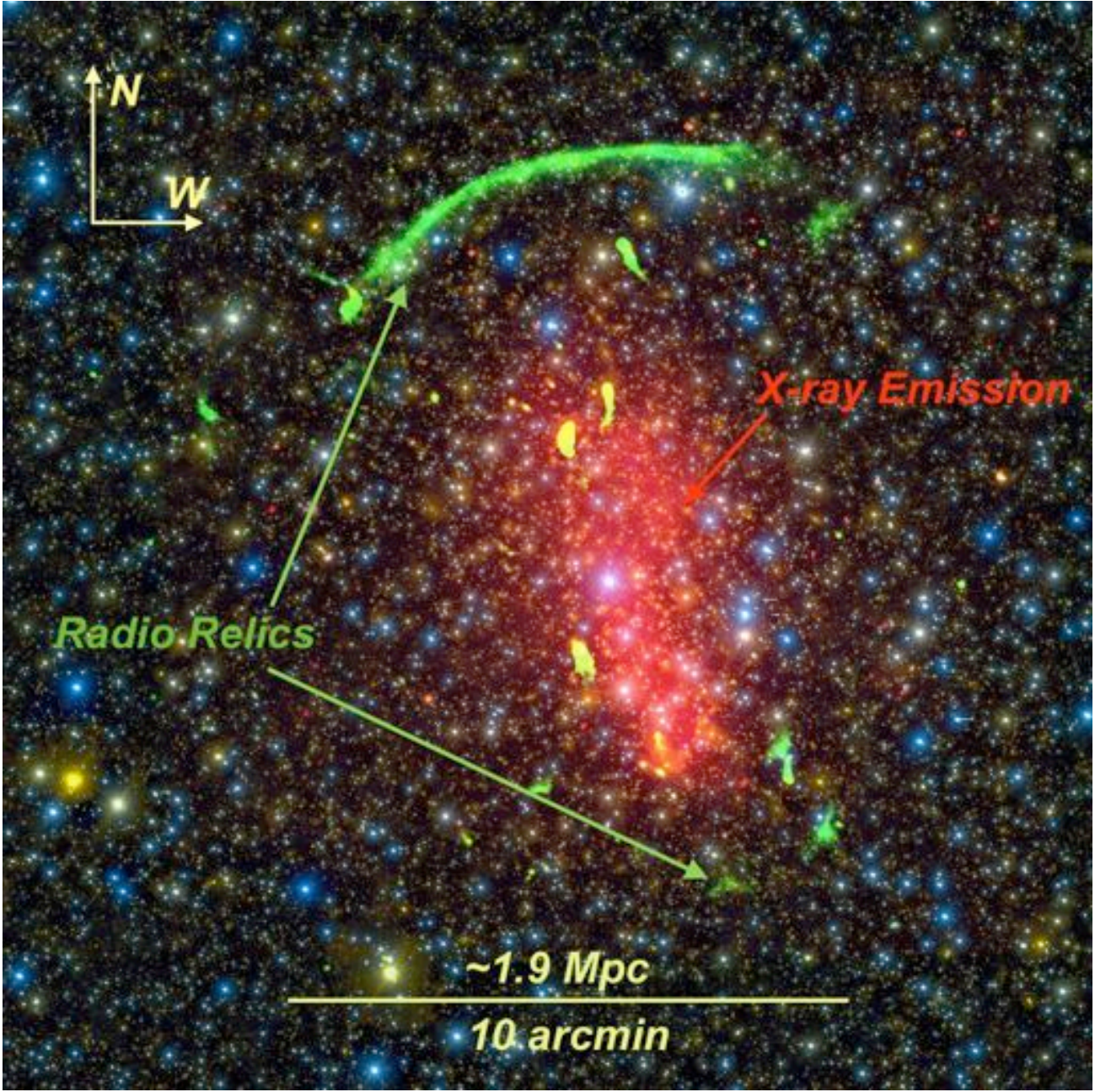}
\vspace{-2cm}
\caption{Illustration of different components in the merging cluster \ciza. The  intensity in green represents the 610 MHz radio emission measured with GMRT (van Weeren et al. 2010). The intensity in red shows the X-ray emission observed with Chandra. The background color-composite is created using Subaru/Suprime Cam data with the $g$, $g+i$, and $i$ filters depicting the intensity in blue, green, and red channels, respectively.   }
\label{fig_ciza_first}
\end{figure*}

This paper is part of a upcoming series of the $MC^2$ publications, presenting our first weak-lensing analysis of \ciza. The cluster is one of the most remarkable clusters in our sample, possessing
a prominent giant radio arc (van Weeren et al. 2010) stretched over $\mytilde2$~Mpc.
Radio relics are discovered at the edge of merging galaxy clusters stretched 
perpendicular to the merger axis and are believed to trace shock fronts (Ensslin et al. 1998).
Therefore, the observed morphology of the radio relics of \ciza ~unambiguously indicates that we are observing a post-merger system colliding along the north-south axis. The north-south elongation of the X-ray emission  from XMM-$Newton$ and $Chandra$ also supports this merger scenario (Ogrean et al. 2013; 2014). In Figure~\ref{fig_ciza_first}, we display the pseudo-color composite image of \ciza~ with
illustration of the radio relics and X-ray emission.

These rich datasets offer opportunities for detailed simulations of the system in order to understand the exact physical mechanism leading to the observed features of \ciza. Our weak-lensing study will provide reliable mass properties of the cluster, which are among the key input parameters of the simulation, but are currently missing. Because of the apparent disruption of the system, we need to derive the mass without relying on any dynamical assumption, which is an important merit of weak lensing probes.

We structure our paper in the following way. In \textsection\ref{section_review}, we briefly review the discovery and development of our understanding of \ciza. \textsection\ref{section_observation} describes the data and reduction method. A basic description of weak lensing theory is presented in \textsection\ref{section_theory}. We present our mass reconstruction  in \textsection\ref{section_mass_reconstruction}.
The cluster mass estimation is described in \textsection\ref{section_mass}.
Our results are discussed in \textsection\ref{section_discussion} before
we conclude in \textsection\ref{section_conclusion}.

We assume a flat $\Lambda$ cold dark matter ($\Lambda$CDM) cosmology with $H_0=70~\mbox{km}~\mbox{s}^{-1}~\mbox{Mpc}^{-1}$, $\Omega_M=0.3$, and $\Omega_{\Lambda}=0.7$. At the redshift of \ciza~ $z\sim0.19$, the plate scale is $\mytilde 3.17~\mbox{kpc}$ per arcsec.

\section{THE GALAXY CLUSTER \ciza: DEVELOPMENT OF UNDERSTANDING } \label{section_review}

The galaxy cluster \ciza~at $(\alpha,\delta)=(22^{h}42^{m}49^{s}.1,~53\degr00\arcmin 51\arcsec)$ or $(l,b)=(104.191\degr,-5.111\degr)$~was discovered in the Clusters In the Zone of Avoidance (CIZA) survey as one of the 57 clusters located at low Galactic latitude (Kocevski et al. 2007). The clusters in the CIZA survey are in general difficult to identify based on galaxy overdensity because of the severe extinction and stellar obscuration. Kocevski et al. (2007) report that the X-ray luminosity from the original ROSAT All Sky Survey (RASS) data is $L_X=6.8\times10^{44}\mbox{erg}\mbox{s}^{-1}$, which is converted to $M_{500}\sim5.5\times10^{14} M_{\sun}$ according to the
scaling relation of Pratt et al. (2009).

The spectacular radio relics of the cluster were discovered by van Weeren et al. (2010) from Westerbork Synthesis Radio Telescope (WSRT; Katgert et al. 1973) and Giant Metrewave Radio Telescope (GMRT; Swarup et al. 1991) data with a clear detection of a spectral index gradient toward the cluster center, which led the authors to conclude that the radio signal is arising from electrons accelerated at large-scale shocks due to a head-on binary merger of roughly equal masses. They also argued that the fraction of the polarization constrains the angle between the plane of the sky and the merger axis to be less than $\mytilde30\degr$. 

Based on the constraints imposed by the radio data, van Weeren et al. (2011) carried out hydrodynamical simulations to determine the merger geometry.
They concluded that  \ciza~ might be undergoing a merger with a mass ratio of 2:1 and an impact parameter less than 400 kpc nearly in the plane of the sky ($<10\degr$). The estimated core passthrough time is about 1 Gyr ago. 
As we will discuss in \textsection\ref{section_discussion}, we believe that the simulation set-up of van Weeren et al. (2011) should be  
revised because the total mass in the two systems there is assumed to be only $5.5\times10^{14} M_{\sun}$, which is based on the estimation of Kocevski et al. (2007), who converted
the ROSAT X-ray luminosity to the mass with the scaling relation of Pratt et al. (2009). Our weak-lensing analysis shows that the total mass of \ciza ~well exceeds $\mytilde2\times10^{15}~M_{\sun}$ (\textsection\ref{section_mass}).

Additional evidence supporting the north-south merger scenario is the presence of tailed radio sources. Stroe et al. (2013) found that 
the tailed radio sources in the \ciza~field are pointing either north or south, consistent with the direction of the merger hypothesis.

The ROSAT Position Sensitive Proportional Counter (PSPC) image already showed that the intracluster medium (ICM) of \ciza~is severely disturbed (van Weeren et al. 2010). A more detailed study of the ICM  was presented by Ogrean et al. (2013) using {\it XMM-Newton} observations. The X-ray emission clearly shows a north-south elongation of the cluster ICM, which is consistent with the merger axis inferred from the radio relics and numerical simulations (van Weeren et al. 2010; 2011). However, no evidence for shock compression is found near the northern relic in the {\it XMM-Newton} data. A weak indication of shock is present near the southern relic, although the feature cannot be exclusively attributed to a shock. Ogrean et al. (2013) concluded that many features in their X-ray data including temperature gradient, surface brightness distribution, etc. suggest that perhaps the actual merger is much more complicated than a simple binary merger, which was the initial hypothesis of van Weeren et al. (2010; 2011).

With the Suzaku X-ray telescope, Akamatsu \& Kawahara (2013) found significant jumps in temperature at the position of the radio relics in \ciza, which is the first confirmation that
the radio relics in the cluster indeed traces the location of the shocks. The implied Mach number from the temperature drop ($\mytilde8$ keV to $\mytilde2$ kev)
is $\mathcal{M}_X=3.15\pm0.52$, 
which is slightly lower than, but statistically consistent with the radio measurement  $\mathcal{M}_{radio}=4.6_{-0.9}^{+1.3}$ derived from the spectral index (van Weeren et al. 2010) . 

Higher resolution X-ray data from Chandra were analyzed by Ogrean et al. (2014). Their investigation of the surface brightness profile does not provide any
evidence of a shock near the northern relic in agreement with their previous study with {\it XMM-Newton} (Ogrean et al. 2013). 
Instead, Ogrean et al. (2014) report that multiple density discontinuities are present in other regions both on and off the hypothesized merger axis, speculating that the X-ray shock features arise, if real, from violent relaxation of the dark matter cores of the clusters. The presence of the large temperature discontinuity of Akamatsu \& Kawahara (2013) was confirmed by
the reanalysis of the Suzaku data by Ogrean et al. (2014).

Stroe et al. (2014a) report that the relic features of \ciza~ are detected even at high-frequency ($\mytilde16$ GHz) using the Arcminute Microkelvin Imager (AMI; Zwart et al. 2008). Interestingly, the high-frequency data detects a clear trend of spectral steepening toward 16 GHz. However,
this finding is somewhat inconsistent with the diffusive shock acceleration model, which is believed to be the main physical mechanism responsible for most radio relics in merging clusters. Another noteworthy feature in the high-frequency data is
a flat-spectrum diffuse extension of the southern relic, which was not present in the lower frequency maps of van Weeren et al. (2010) and Stroe et al. (2013).

The H$\alpha$ survey of the \ciza~ field (Stroe et al. 2014b) shows  a significant excess of $H\alpha$ sources with respect to the blank field.
Possibly, this excess is linked to the enhanced star formation activities triggered by the passage of the shock wave.

\begin{figure*}
\includegraphics[width=9cm]{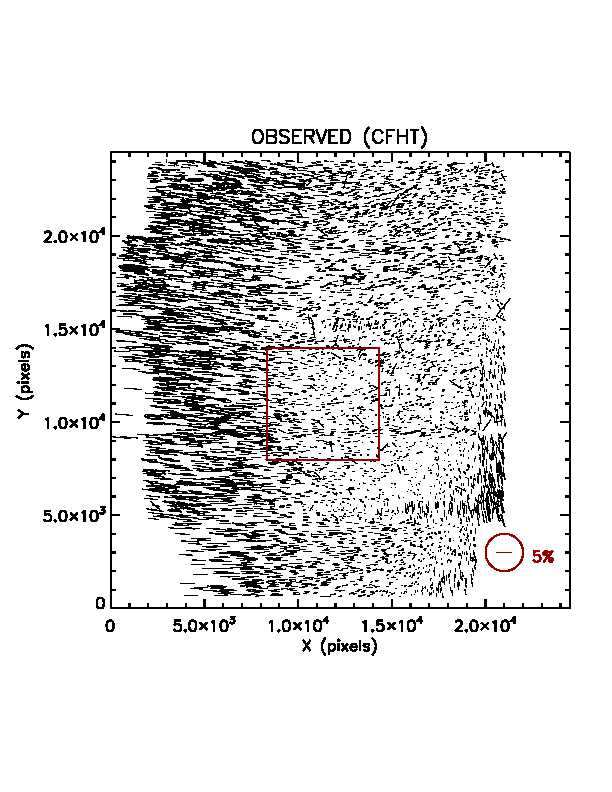}
\includegraphics[width=9cm]{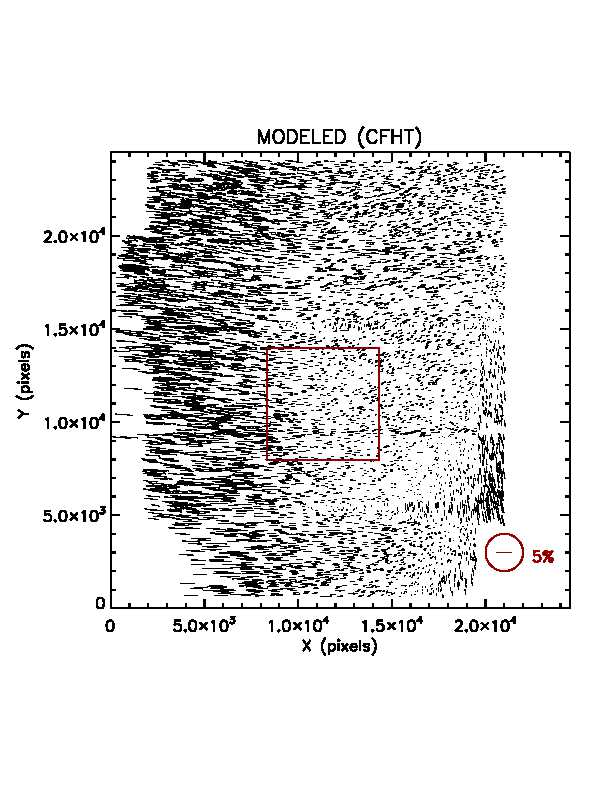}
\includegraphics[width=9cm]{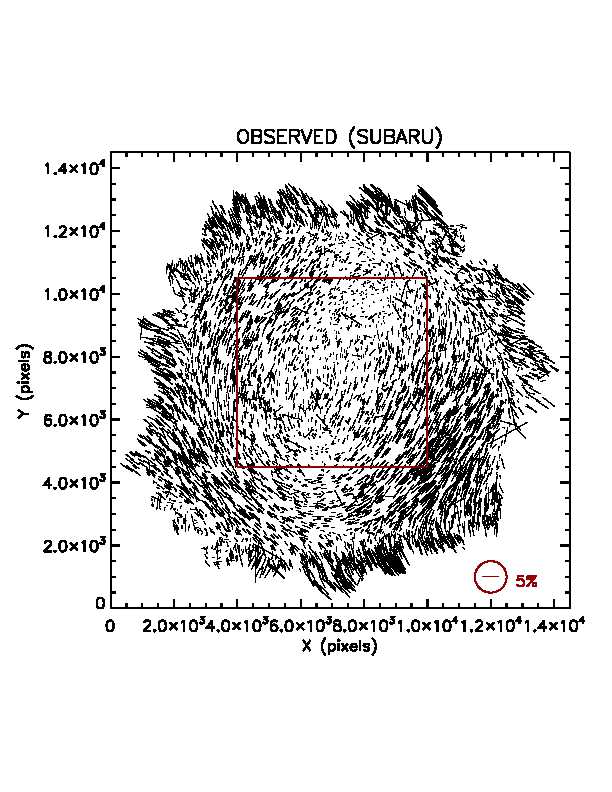}
\includegraphics[width=9cm]{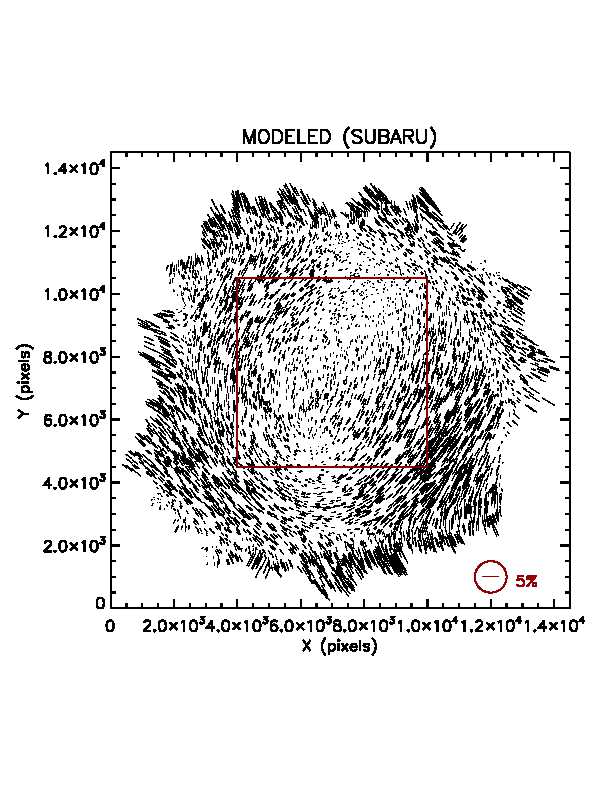}
\vspace{-2cm}
\caption{Comparison of ellipticities between observed stars and model PSFs. In the top panel, we display the comparison in CFHT/MegaCam data ($r$ band). 
We do not remove spurious ellipticities in the left panel, which are contributed by severely blended stars or galaxies mis-identified as stars.
Except for these outliers, the agreement is excellent. This agreement also serves as a verification that no large systematic error is present in image registration, in which case we would detect systematic elongation of stars only in the left panel (the model in the right panel assumes that image alignment is perfect).
The red box indicates the central $20\arcmin\times20\arcmin$ region, where we perform our two-dimensional mass reconstruction. The encircled red stick on the lower right corner shows the size of 5\% ellipticity. In the bottom panel, the result is for the Subaru/SuprimeCam data in the $i$ band.}
\label{fig_psf_reconstruction}
\end{figure*}

\section{OBSERVATIONS} \label{section_observation}

\subsection{Subaru/SuprimeCam Data Reduction}
\ciza ~was observed with Subaru/Suprime Cam on 13 July 2013 in $g$ and $i$ with integrations of 720~s and 3,313~s (PI. D. Wittman). We rotated the field of each visit  in order to distribute the bleeding trails and diffraction spikes from bright stars azimuthally and later removed them by median-stacking different visits. This scheme enables us to maximize the number of galaxies usable for lensing. The median seeing for $g$ and $i$ are $0\farcs72$ and $0\farcs65$, respectively.

We used the SDFRED (Ouchi et al. 2004; Yagi et al. 2002) package to subtract over-scan and bias, make flats, correct for geometric distortion (3rd order polynomial), and mask out regions affected by bad pixels and auto guide (AG) probe. Then, we removed residual distortion and refined the astrometric solution with the SCAMP software (Bertin 2006)\footnote{http://www.astromatic.net/software/scamp}, which compares the coordinates of the common objects in individual frames with those in a reference catalog. From our experience, we find that the Sloan Digital Sky Survey Data Release 7 (SDSS DR7; Abazajian et al. 2009) reference catalog in general provides the most reliable astrometric solution if the field is observed with SDSS. However, the \ciza  ~field is not within the SDSS survey. Nevertheless, we were still able to achieve weak-lensing quality accuracy (rms$\mytilde0.02$ pixel) in image registration utilizing the Two Micron All Sky Survey (2MASS; Skrutskie et al. 2006) catalog. 

Finally, we created a large mosaic $\mytilde48\arcmin\times46\arcmin$ image with the SWARP package (Bertin et al. 2002)\footnote{http://www.astromatic.net/software/swarp} in two passes. In the first pass, we generated a median-stack image. In the second pass, we created a coadd stack by weight-averaging input frames. Because it is necessary to identify the pixels that should be excluded in co-addition, we masked  them out by modifying the weight files of input frames after identifying them by comparing pixel values of input frames with those of the median stack; the current version (v2.19.1) of SWARP does not automate this procedure. 

\subsection{CFHT/MegaCam Data Reduction}
The cluster image was taken with CFHT/MegaCam during the 3-12 July 2013 period in $r$ (PI. A. Stroe). The total integration is 24,000~s, consisting of 40 short (600~s) exposures. The median seeing is $\mytilde0\farcs74$.
The field rotation mentioned above for the Subaru observation was not implemented because of the technical restriction of the instrument. Raw level data reduction was carried out with the Elixir pipeline\footnote{http://www.cfht.hawaii.edu/Instruments/Elixir/}. However, similarly to the Subaru data, we used a combination of SCAMP and SWARP to generate our final mosaic.

\subsection{Object Detection, Photometry, and Extinction Correction}
We run SExtractor (Bertin \& Arnouts 1996) in a dual image mode using the $i$-band image for detection.
Prior to object detection, we choose to let SExractor convolve 
the $i$-band image with a Gaussian kernel whose size approximately matches the seeing.
Objects are identified by looking for at least five connected pixels above two times the sky rms.
The blending threshold parameter ({\tt BLEND\_NTHRESH}) is set to 64 with a minimal contrast of {\tt DEBLEND\_MINCONT}$=10^{-4}$.
We employ isophotal magnitudes ({\tt MAG\_ISO}) to estimate object colors whereas total magnitude ({\tt MAG\_AUTO}) is used to
compute luminosity.

Given the low Galactic latitude of the cluster, the measured magnitudes are highly attenuated by dust extinction, which varies across the field (e.g. from $\mytilde 0.6$ to $\mytilde 0.9$ in the $i$-band). In order to correct for this effect and recover intrinsic magnitudes, we used the redenning values from Schlafly \& Finkbeiner (2011), with a spatial resolution of 
$\mytilde 4\arcmin$. We interpolate between the extinction pierce points using cubic interpolation to predict the dust attenuation at each source position. We correct the $g$, $r$ and $i$ magnitudes by interpolating in wavelength to the effective wavelength of the Subaru and CFHT filters (see Stroe et al. 2014b for details).
The 5$\sigma$ limiting magnitudes are 24.9, 24.1, and 25.4 for the $g$, $r$, and $i$ filters, respectively.

\subsection{Point Spread Function Modeling}

A point spread function (PSF) distorts galaxy shapes, and thus it is important to carefully model it  and remove its effect when we measure
weak lensing signals.
Our PSF modeling is based on the principal component analysis (PCA) described in-depth in Jee et al. (2007) and Jee \& Tyson (2010). The method has been successfully applied to a wide range of data from ground (e.g., Jee et al. 2013; Dawson et al. 2012) to space (e.g., Jee et al. 2014). In the most recent competition called the 3rd GRavitational lEnsing Accuracy
Testing (GREAT3; Mandelbaum et al. 2014), the method (team name: sFIT) won the data challenge.
The application to ground-based data was described in detail in Jee et al. (2013). Below, we briefly summarize the key steps.

The SWARP software allows us to save the intermediate files (RESAMP) to be used for the final co-add. These files are already resampled to match the final co-add exactly (only integer pixel shifts are required). Our PSF model starts with identifying stars from these intermediate resampled images.
In the \ciza~field, the typical number of stars per CCD usable for PSF modeling is $\mytilde 400$ or higher. This is more than enough to obtain reliable principal components (we kept the most significant 20 principal components) and describe the coefficients along these principal components using a polynomial interpolation method. Although recent studies suggest that a more sophisticated PSF interpolation method is required for future cosmic shear surveys\footnote{http://www.great3challenge.info}, we find that a simple 3rd order polynomial interpolation method provides sufficient accuracy in the current cluster weak-lensing analysis.
Because we choose to measure shapes from final co-adds, we model the PSF in the co-adds by propagating the PSF models of individual CCD frames.

The fidelity of the result obtained from this procedure can be examined by comparing the PSF statistics directly measured from stars in the co-add with those inferred from our PSF model propagated from individual exposure data.
In  Figure~\ref{fig_psf_reconstruction}, we present such comparisons for both Subaru and CFHT images. A few features are noteworthy here. First, the PSF pattern in the coadd is complicated and shows a large spatial variation. Second, conspicuous discontinuities in the PSF variation is present 
(often these features are found across the CCD boundaries). Third, the PSF pattern predicted by our PCA-based model closely matches the observed PSF pattern. This agreement verifies that the PSF model obtained after stacking PSFs from individual exposures is indeed valid in representing the PSF in the co-add. In addition, it also warrants that there is no systematic error in our image registration. For example, if there is any frame misaligned by as few as $\mytilde0.5$ pixels in the $x$-axis, the (observed) stars in that region would possess large ellipticities along the $x$-axis in the coadd image, which in turn would cause large discrepancies between observation and model.

\subsection{Shape Measurement} \label{section_shape}

We use forward-modeling to determine galaxy ellipticity. The details are described in our past publications (e.g., Jee et al. 2013).
In brief, we first convolve elliptical gaussian with a PSF profile and then fit the result to galaxy images. We define the ellipticity with
\begin{equation}
e=\frac{a-b}{a+b},
\end{equation}
\noindent
where $a$ and $b$ are the semi-major and -minor, axes, respectively, that are determined from the fit. Because real galaxy profiles are not Gaussian, the result is biased. Also, photon noise causes non-negligible bias in the shape parameters, in particular, for faint objects.
In Jee et al. (2013), we determined the correction factor by carrying out image simulations with real galaxy images and found that our ellipticity should by multiplied by $\mytilde1.1$ to recover input shears. This procedure calibrates out the isotropic part of the shear bias. Typically, shear measurement is also subject to anisotropic bias often correlated with PSF elongation.
We corrected for this additive bias in our cosmic shear study presented in Jee et al. (2013). However, this additive bias is several orders of magnitude smaller than the statistical error and thus can be safely ignored in the current cluster lensing study. 

As the ellipticity has both direction and magnitude, it is convenient to represent it using the following two components:
\begin{equation}
e_+= e \cos(2 \theta)  \label{eqn_e1}
\end{equation}
\begin{equation}
e_\times= e \sin(2 \theta), \label{eqn_e2}
\end{equation}
\noindent
where $\theta$ is the angle between the $x$-axis and the major axis of the ellipse.

\subsection{Multiplicative Bias by Stellar Contamination} \label{section_stellar_contamination}

Isolated stars with a reasonably high S/N are easily identified by their shape parameters
and reliably removed from source catalogs.
However, inevitably, some stars are also blended with other stars or galaxies. These blended objects
when included into a source catalog as single objects dilute lensing signals because stars do not contain any lensing signal.
Hoekstra et al. (2014) estimated this dilution effect using the stellar population model of the Galaxy by Robin et al. (2003) and the lensing
image simulaton tool GalSim\footnote{http://https://github.com/GalSim-developers/GalSim}.
They report that at $\left | b \right | \gtrsim40\degr$ the dilution effect by star contamination is much less than $\mytilde1$\%. However, as the
Galactic latitude decreases below $\left | b \right | <40\degr$, the dilution effect is rapidly increasing, reaching $\mytilde2$\% 
at $b=20\degr$.  Because the simulation of Hoekstra et al. (2014) did not cover the range that includes our case ($\left | b \right | \sim5\degr$), we have to extrapolate their result to estimate the dilution effect at $\left | b \right | =5\degr$, which gives $\mytilde5$\% compared to the case at $b=90\degr$.
We apply this correction in addition to our shear calibration mentioned in \textsection\ref{section_shape}. However, this dilution effect is still small relative to the shot noise, which already causes as large as $\mytilde20$\% error in cluster mass.

\subsection{Source Selection} \label{section_source_selection}

\begin{figure}
\includegraphics[width=8cm]{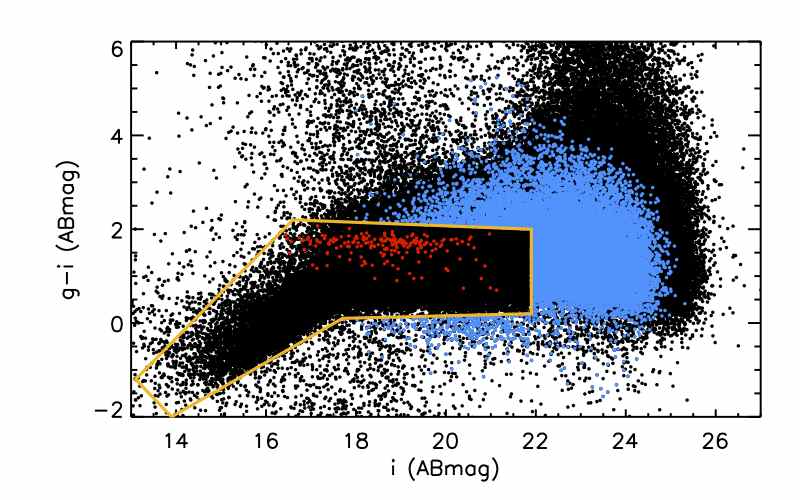}
\caption{Source selection based on color-magnitude relation in the \ciza~field. Extinctions are corrected for. The yellow polygon encloses an area, which is mostly populated with stars and the cluster members. The red dots represent the 218 spectroscopically confirmed members of \ciza. The sources selected for weak-lensing analysis are shown in blue.}
\label{fig_cmd}
\end{figure}

We base our source selection mainly on the color-magnitude relation with some additional steps to minimize the contamination from stars misidentified as galaxies. Figure~\ref{fig_cmd} shows the relation between $g-i$ and $i$. The objects in the yellow polygon were explicitly rejected because most objects there are either stars or cluster members. The blue dots represent the sources that we select for weak-lensing analysis. 
We do not use the $r$-band photometry in this selection because photometric redshift estimation with the three $g$, $r$, and $i$ bands does not provide significant advantage in isolating the cluster members. In addition, it is difficult to obtain reliable $r$-band photometry for many sources because of the charge bleeding.

In addition to the above color-magnitude relation, we require the object half light radius (in the $i$ band) to be greater than $0\farcs44$, which is the upper limit of the stellar locus in the current data. Also, the pre-seeing semi-minor axis (determined after PSF-convolved elliptical Gaussian fitting) should be significantly greater than zero. This last measure is needed to remove remaining stars and avoid the aliasing effect pertinent to small objects.
We also applied S/N cuts by ensuring that the ellipticity error per component is less than 0.3 and the detection significance is above 5 $\sigma$.  From the Subaru $i$-band image, we obtain $\mytilde~16$ galaxies per sq. arcmin, which is roughly a factor of two lower than the typical number that we achieve with Subaru in other fields (e.g., Dawson et al. 2012). We repeat that this low source density is due to both high stellar density and high extinction. Because of both relatively shallow depth (the 5~$\sigma$ limiting mag is $r_{limit}\sim24.1$) and obscuration by bleeding trails, the source number density is a factor of two lower ($\mytilde8$ galaxies per sq. arcmin)  in the CFHT $r$-band data.
Although we merge the two shape catalogs from Subaru and CFHT to perform our final weak-lensing analysis, the result is consistent with the case when only Subaru shapes are used. This is not surprising because most of the objects detected in the CFHT image are also visible in the Subaru image. However, it is important to note that the results based on the CFHT shapes are also statistically consistent with the results obtained with the Subaru shapes. In particular, we emphasize that the two versions of mass maps from both Subaru and CFHT are in excellent agreement, which will be demonstrated in \textsection~\ref{section_mass_reconstruction}.

\subsection{Cluster Member Selection} \label{section_member_selection}

Detailed comparison of lensing results with cluster galaxies provides useful probes to the interplay between different constituents of galaxy clusters.
To facilitate comparisons between the mass and member galaxy distributions in \ciza, we select a separate sample of likely cluster members as follows.

We use the color-magnitude relation of the red cluster members in conjunction with our spectroscopic survey data of the cluster field.
The locus of the red-sequence is determined using the 217 spectroscopic members. Although this spectroscopic sample
contains a significant number of blue members, the tight relation (Figure~\ref{fig_cmd}) allows us to define a
narrow strip ($\delta (g-i)\sim0.6$) in the magnitude range $16.5<i<20$ to isolate the red-sequence galaxies.
Stars are removed from this initial selection by discarding objects if
they are spectroscopically confirmed stars, their half light radius is greater than $r_h= 0\farcs44$, or their semi-minor axis
after deconvolution is statistically consistent with zero.
The cluster member catalog is further refined by visually inspecting the object to reject any conspicuously blended stars.
Finally, we add the blue spectroscopic members that are located outside the initial narrow strip.
We compare the smoothed luminosity and number density maps with our mass reconstruction in \textsection\ref{section_mass_reconstruction}.

\section{BASIC WEAK-LENSING THEORY AND SOURCE REDSHIFT ESTIMATION} \label{section_theory}

Interested readers are referred to Bartelmann \& Schneider (2001) and Hoekstra (2013) for more details on weak lensing and its application
to galaxy clusters. Here we only provide a brief review.

In a  weak-lensing regime, we can assume that averaging the ellipticity components defined in Equations~\ref{eqn_e1} and~\ref{eqn_e2} become reduced shears $g_1$ and $g_2$, respectively. That is,
\begin{equation}
\left < e_+ \right > \simeq g_1  \label{eqn_ep}
\end{equation}
and
\begin{equation}
\left < e_{\times} \right > \simeq g_2. \label{eqn_em}
\end{equation}
The reduced shears  comprise the matrix elements of the following weak-lensing transformation:
\begin{equation}
\textbf{A}=(1-\kappa) \left (\begin{array} {c c} 1 - g _1 & -g _2 \\
                      -g_2 & 1+g _1
          \end{array}  \right ), \label{eqn_lens_trans}
\end{equation}
\noindent
where $\kappa$ is the surface mass density in units of the critical density:
\begin{equation}
\Sigma_c = \frac{c^2}{4 \pi G D_l \beta } \label{eqn_sigma_c}.
\end{equation} 
In Equation~\ref{eqn_sigma_c}, $c$ is the speed of light, $G$ is the gravitational constant,
and $D_l$ is the angular diameter distance to the lens. $\beta$ is defined below.

The lensing signal strength depends on the distance ratios between lens, source, and the observer. In order to interpret and quantify
the observed signal properly, it is necessary to estimate the source redshift. Often, a lensing efficiency
is expressed in terms of $\beta$ defined as
\begin{equation}
\beta=max \left [ D_{ls}/D_s,0 \right ] 
\end{equation}
where $D_{ls}$ and $D_s$ are the angular diameter
distances between the lens and the source, and between the observer
and the source, respectively. Because sources in front of the lens do not contribute to the lensing signal, we assign zero to $\beta$
in those cases ($D_{ls}/D_s$ becomes negative).
Therefore, we need to obtain $\left < \beta \right >$ to characterize the effective lensing efficiency of the source population. We refer to the corresponding redshift as effective redshift $z_{eff}$. However, because the lensing efficiency is non-linear, 
using $\left < \beta \right >$ alone to represent a broad distribution
biases the result. A first order correction is given when we update the reduced shear $g$ with the following equation (Seitz \& Schneider 1997):
\begin{equation}
g^{\prime} = [1 + (\left
  <\beta^2 \right >/ \left <\beta \right >^2-1) \kappa ]g, \label{eqn_beta2}
\end{equation}
where $\kappa$ is the surface mass density in units of the critical surface density $\Sigma_c$ (see \textsection\ref{section_theory} for the definition of $\Sigma_c$).

Since we base our selection on the two-filter broadband photometry as described in \textsection\ref{section_source_selection}, we estimate the source redshift indirectly utilizing a publicly available reference catalog. We choose the Cosmic Evolution Survey (COSMOS) photometric
redshift catalog\footnote{available at http://irsa.ipac.caltech.edu.} (Ilbert et al. 2009) as our reference because 1) the magnitude-limit $I<25$ is well-suited to the current analysis, 
2) the result is obtained based on 30 filters, and 3) the survey area is reasonably large (2 sq. degs) so that the result does not greatly suffer from the
cosmic variance. In addition, the Ilbert et al. (2009) photomery is also derived from the Subaru Telescope, which obviates the need for any
transformation between the current \ciza  ~and the COSMOS photometric systems.

We apply the source selection criteria described in \textsection\ref{section_source_selection}
to the COSMOS galaxy catalog. We obtain $\left < \beta \right >=0.656$ and
the corresponding effective redshift of this subset is $z_{eff}=0.626$.
Because the depth of our \ciza ~image is shallower than the COSMOS images, it is necessary
to correct for the difference. We compute this correction factor by constructing
magnitude histograms for the two catalogs and weight galaxies according to the ratio of our source number
density per magnitude bin to the COSMOS one. The stellar contamination discussed in \textsection\ref{section_stellar_contamination} is also taken into account in this step.
Both $\left < \beta \right >$ and $z_{eff}$ decrease to 0.616 and 0.549, respectively.
We obtain $\left <\beta^2 \right >$=0.438, which is needed to account for
the width of the redshift distribution (Equation~\ref{eqn_beta2}).

\section{TWO DIMENSIONAL MASS RECONSTRUCTION} \label{section_mass_reconstruction}

The reduced shear defined in Equations~\ref{eqn_ep} and \ref{eqn_em} is directly observable (up to
shear calibration) by averaging object ellipticities. However, in order to obtain the surface mass
density $\kappa$, we need to know shears $\gamma$, which are related to reduced shears via:
\begin{equation}
g=\frac{\gamma}{1-\kappa}.
\end{equation}
Often, $g=\gamma$ is assumed
in the very weak-lensing regime where $\kappa \ll 1$.  Under this assumption, we can obtain the two-dimensional convergence field by performing the following integral:
\begin{equation}
\kappa (\bvec{x}) = \frac{1}{\pi} \int D^*(\bvec{x}-\bvec{x}^\prime) \gamma (\bvec{x}^\prime) d^2 \bvec{x}, \label{k_of_gamma}
\end{equation}
\noindent
where $D^*$ is the  convolution kernel defined as $D(\bvec{x} ) = - 1/ (x_1 - i x_2 )^2$ and $\bvec{x}$ is the coordinate.
This direct inversion first used in Kaiser \& Squires (1993) is still a popular method to reconstruct a two-dimensional mass distribution. 

However, near cluster centers, $\kappa$ is non-negligible, and thus we need to include the non-linear relation between $g$ and $\gamma$. In the current paper, we used the inversion code of Jee et al. (2007b), which utilizes the entropy of the $\kappa$ field to regularize the result.

\begin{figure*}
\includegraphics[width=9cm]{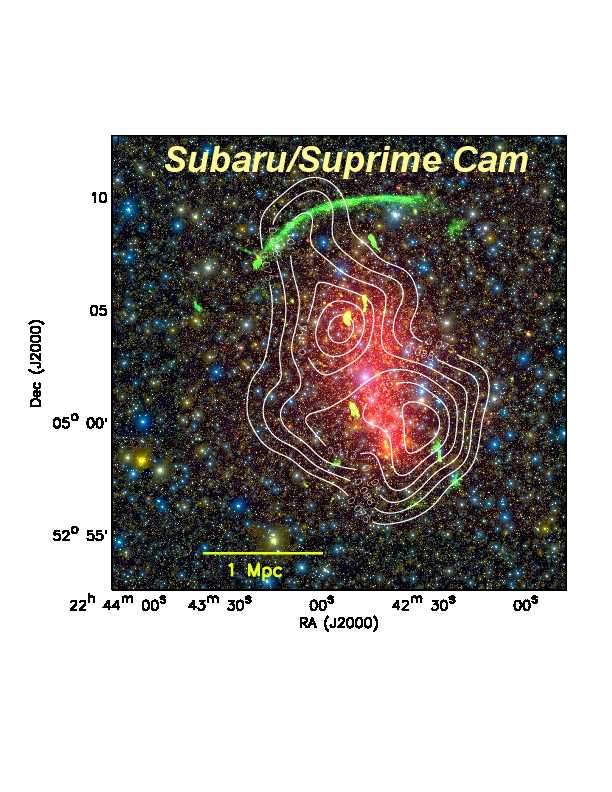}
\includegraphics[width=9cm]{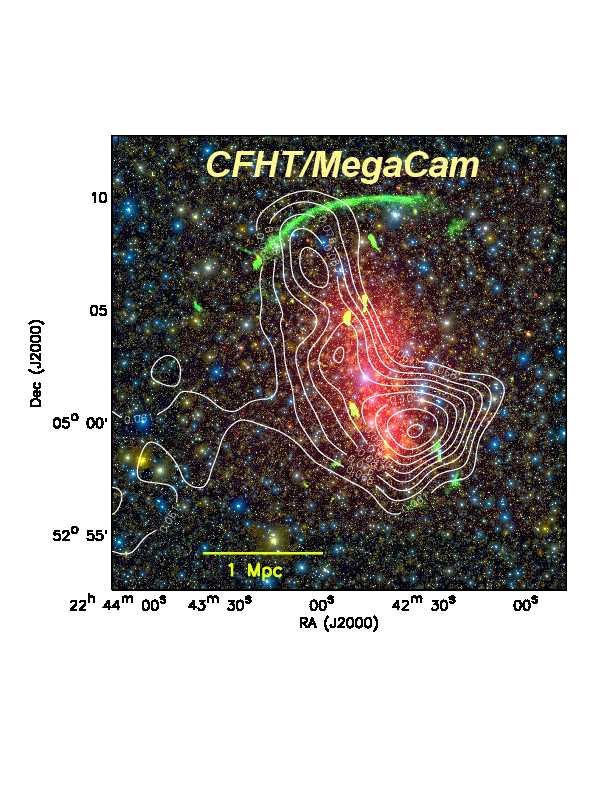}
\vspace{-2cm}
\caption{Weak-lensing mass reconstructions from Subaru (left) and CFHT (right). The white contours represent the projected mass density ($\kappa$), which
is subject to the mass-sheet degeneracy ($\kappa \rightarrow \lambda \kappa + 1-\lambda$).
We arbitrarily scale the mass map in such a way that the average $\kappa$ value becomes zero near the field boundary. The Subaru and CFHT results are
consistent with other each. The consistency between independent telescopes serves as verification that the observed substructures in both maps are not due to residual systematics. Readers are reminded that the PSF patterns are very different between the two observations (Figure~\ref{fig_psf_reconstruction}). It is also important to note that the source number density from the CFHT data is a factor of two lower. 
}
\label{fig_2D_mass}
\end{figure*}

We show our mass reconstruction results in Figure~\ref{fig_2D_mass} from both Subaru and CFHT. Clearly, the results demonstrate that the mass distribution is elongated along the merger axis inferred from the radio relics. Note that the substructures seen in both instruments are consistent.
We regard this consistency as verification that the observed substructures are not due to any residual systematics in weak-lensing measurements.
As shown in Figure~\ref{fig_psf_reconstruction}, the PSF ellipticity patterns of Subaru and CFHT are different. Thus, if our PSF correction error were significant, we would not observe this level of consistency between the two instruments. 
Also, remember that the source number density in the CFHT weak-lensing catalog is a factor of two lower. This implies that the CFHT result
can be regarded as one of the bootstrapping realizations of the Subaru result (in the absence of systematic errors).
Most of the sources used for the CFHT weak-lensing analysis are present in the Subaru source catalog. Therefore, little difference from the Subaru result is observed when we combine the two source catalogs.
Nevertheless, we note that our weak-lensing analysis hereafter is based on the union of both Subaru and CFHT shape catalogs.
When objects appear both in Subaru and CFHT, we combine their shapes by weight-averaging their ellipticities.
Weights are computed using ellipticity errors, which are in turn derived from a Hessian matrix. 

\begin{figure*}
\includegraphics[width=9cm]{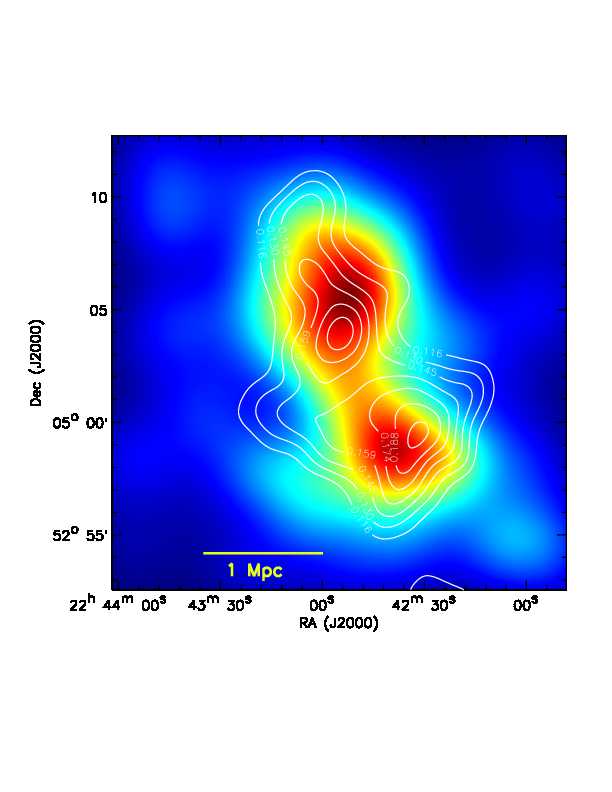}
\includegraphics[width=9cm]{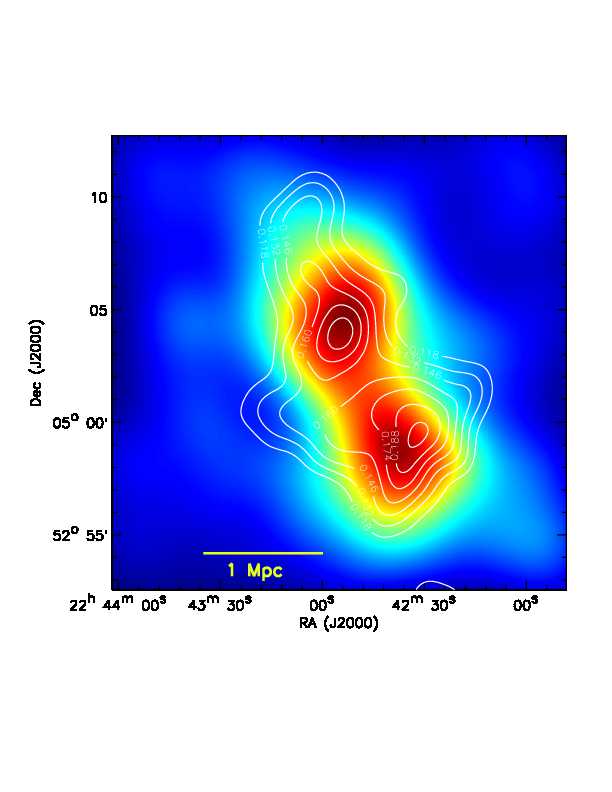}
\vspace{-2cm}

\caption{Mass distribution overlaid on
galaxy distribution. The white contours represent the mass reconstruction when we combine shapes from CFHT and Subaru.
The background is color-coded with smoothed galaxy luminosity (left) and number density (right).
Smoothing is done with a Gaussian with a FWHM$\sim 3 \farcm4$.
}
\label{fig_massovergal}
\end{figure*}

Figure~\ref{fig_massovergal} compares the mass reconstruction from the combined catalog (CFHT+Subaru) 
with the cluster galaxies. 
The cluster member selection is described in \textsection\ref{section_member_selection}. We
smooth the galaxy distribution using a Gaussian kernel with a FWHM$\simeq3\farcm4$.
Both number and luminosity maps indicate that the cluster galaxies in \ciza~ have a bimodal distribution, and our mass reconstruction reveals two dominant mass clumps that can be associated with the two peaks in both luminosity and number density maps.
It is apparent that the mass centroids do not perfectly align with the cluster galaxies. The mean offset between the mass and luminosity/number density peaks is $\mytilde1\arcmin$. We present our analysis on the statistical significance of the mass-galaxy offset in 
\textsection\ref{section_mass_galaxy_offset}.

\section{HOW MASSIVE IS \ciza?} \label{section_mass}

Together with the radio relics, galaxy distribution, and X-ray morphology, our weak-lensing mass reconstruction shows that \ciza~is comprised of at least two massive halos separated by $\mytilde1$~Mpc.  This is analogous to the situation where we 
need to perform accurate photometry for two blended objects that are of comparable luminosity.
Ignoring the presence of the companion will lead to significant overestimation of each object's luminosity. 
Of course, this companion bias is avoided when we fit the two objects simultaneously and estimate the luminosity based on the resulting 
best-fit parameters. Similarly, if we assume a single halo and attempt to estimate the mass using the azimuthally averaged shear  profile, the resulting mass will become non-negligibly biased. Hence, for the mass estimation of \ciza, we fit the two
halos simultaneously. One caveat in this approach is that we still have to assume that both halos can be represented by a Navarro-Frenk-White (NFW; Navarro et al. 1997 ) profile. In general, the difference between our model and actual cluster profiles is an important topic for cluster cosmology, where one desires to know the mass with bias smaller than 1\% accuracy. In this study, we do not suggest a solution to this cluster
model bias. Instead,  we perform a separate mass
estimation using aperture mass densitometry (Fahlman et al. 1994; Clowe et al. 2000), which provide non-parametric masses within an aperture, and compare these projected masses with the results derived from our parametrization. We demonstrate that the discrepancy is negligible within the current statistical error. 

\begin{figure}
\includegraphics[width=9cm]{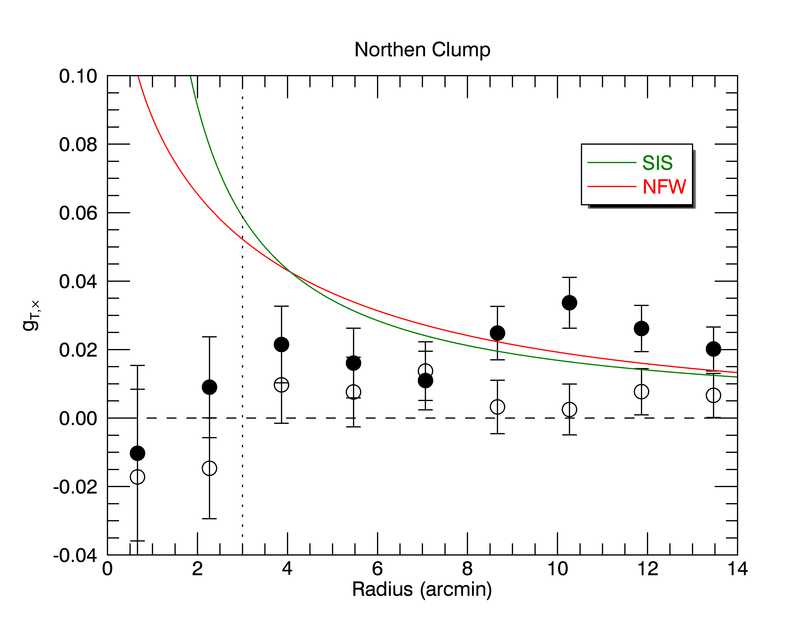}
\includegraphics[width=9cm]{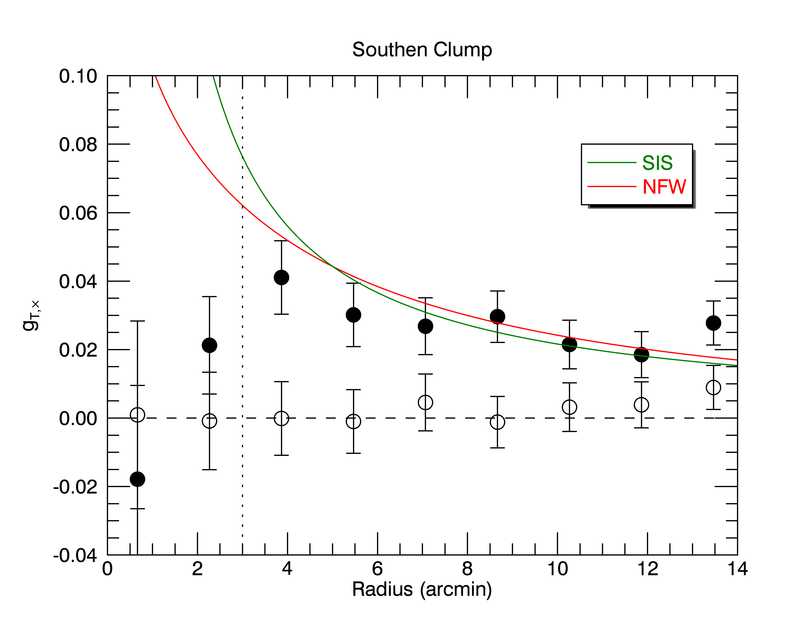}
\caption{Tangential shear profile of \ciza. The top and bottom panels show the tangential shear profile for the northern and southern components, respectively. We choose the luminosity centers (Figure~\ref{fig_massovergal}) as references for both. The filled circles represent the weak-lensing signal whereas the open circles show so-called B-mode signal, which should be consistent with zero as observed if the lensing signal comes from pure E-mode. 
We refer to Table~1 for the profile fitting results. The vertical dotted line is to denote the minimal radius inside which we exclude the data for fitting.}
\label{fig_tan_shear}
\end{figure}

Although we emphasize above that the traditional method of fitting
an analytic profile to one-dimensional (1D) tangential shear profile causes bias, we still present the results from this approach in Figure~\ref{fig_tan_shear} because the results give rough estimates and provide opportunity to examine a possible presence of systematics (so-called B-mode test). The B-mode signal is consistent with zero, which shows that no significant systematic errors are present in our analysis. The Singular Isothermal Sphere (SIS) model is used to predict velocity dispersions whereas the NFW model is assumed for
mass estimation.
We summarize the fitting results in Table 1, which are to be compared to the results from our simultaneous fitting explained below.

Because we desire to determine parameter uncertainties and examine their correlations robustly, we  perform Markov-Chain-Monte-Carlo (MCMC) analysis. In fact, this approach was also used in the case of \elgordo~  (Jee et al. 2014b), which closely resembles \ciza~ in terms of their structure and mass.
\elgordo, nicknamed ``El Gordo" (Menanteau et al. 2012) meaning ``the Fat One" in Spanish, is a merging system with two comparably massive ($\mytilde10^{15}M_{\sun}$ each) clumps.
Similarly to the case of \elgordo, we place two NFW halos over the northern and southern sub-clusters. For each NFW halo, the Duffy et al. (2008) mass-concentration relation is assumed. We choose the galaxy luminosity center as an initial centroid and allow it to vary as we run our MCMC. We use a flat prior with a bounding box $200\arcsec\times200\arcsec$ centered on the luminosity peak.  
For the concentration parameter, we also use a flat prior ranging from c=2 to 4 for both halos with the 1D fitting results as initial values. 
We exclude  source galaxies when they are located within $r_{cut}=200\arcsec$ circle. The mass varies $\mytilde11$\% when $r_{cut}$  varies between $150\arcsec<r_{cut} < 250\arcsec$.
 
\begin{figure}
\includegraphics[width=9cm]{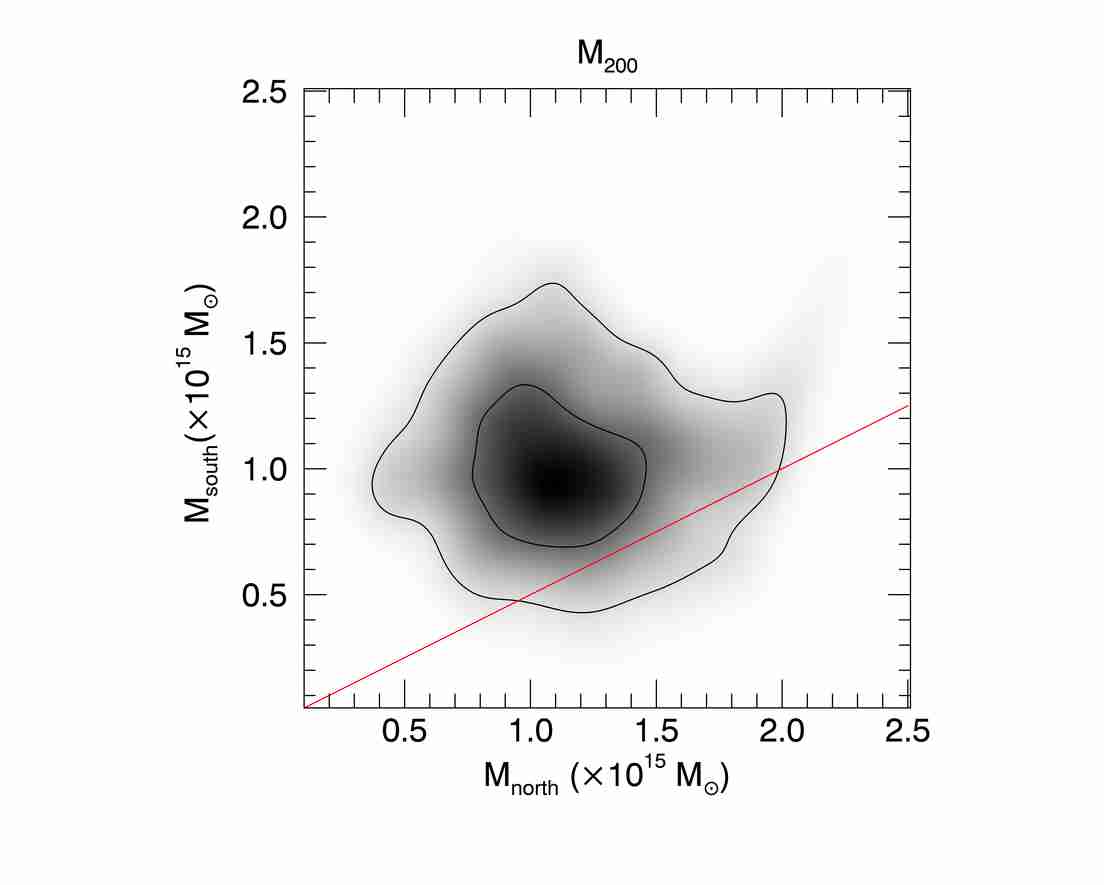}
\caption{Mass constraints on \ciza~ from simultaneous 2D fitting. The inner and outer contours correspond to 1 and 2 $\sigma$ confidence.
The solid red line depicts the 2:1 mass ratio assumed in the simulation of van Weeren et al. (2011).
 }
\label{fig_c_and_m}
\end{figure}

The resulting parameter constraints are displayed in Figure~\ref{fig_c_and_m}. 
The most probable masses are $M_{200c}\sim10^{15}M_{\sun}$ for both clusters. We summarize the fitting results in
Table 2. The mass of the southern halo based on the 1D analysis is higher by $\mytilde60\%$, although the results are in good agreement for the northern clump.
The large difference in the southern subcluster indicates that 
cluster mass estimation with weak-lensing requires care when non-negligible substructures are present.  

\begin{figure}
\includegraphics[width=9cm]{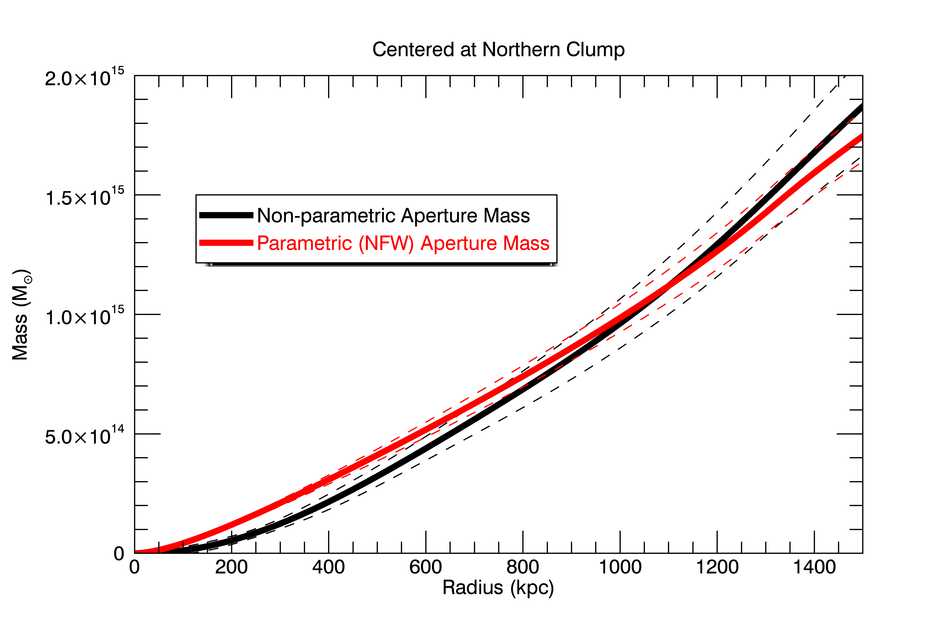}
\includegraphics[width=9cm]{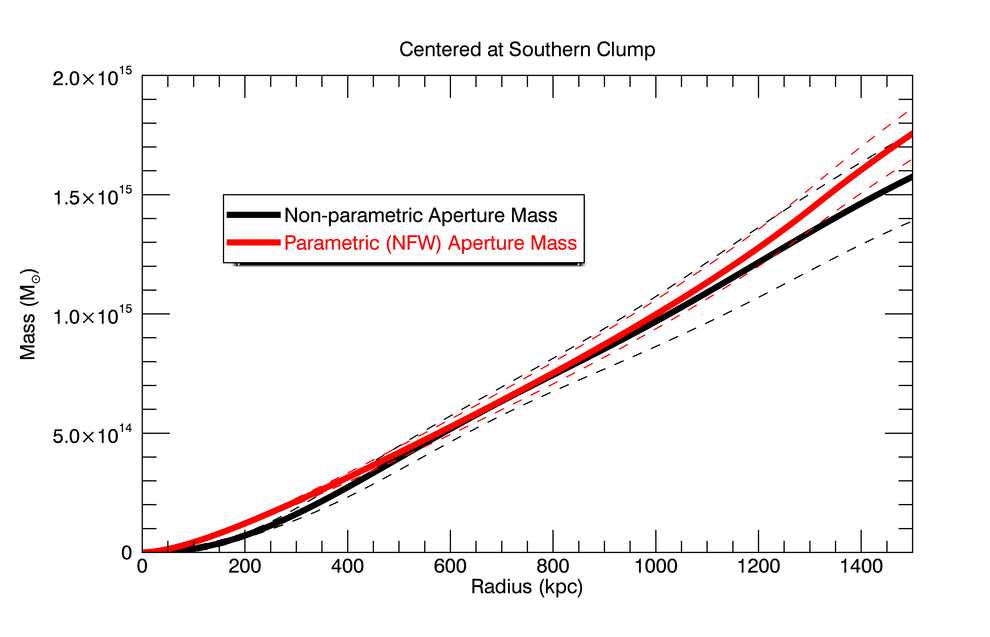}
\caption{Comparison of projected masses between aperture mass densitometry and NFW fitting. The aperture mass densitometry is performed using the luminosity center as a origin. We set up a control annulus at $r=1.9-2.5~\mbox{Mpc}$. The density within this region is estimated utilizing the two-dimensional NFW fitting results.
For computation of the aperture mass based on NFW profile fitting, we stack two projected NFW profiles created with the parameters shown in Table 1. The results from the two different methods are in excellent agreement, indicating that despite the apparent violent merger, parametrizing the halos with NFW profile is still a good approximation in \ciza. The dashed lines depict 1-$\sigma$ (lower and upper) limits. 
 }
\label{fig_aperture_mass}
\end{figure}

The aperture mass densitometry can be performed by evaluating the following integral:
\begin{eqnarray}
\zeta _c (r_1, r_2,r_{max})  =  \bar{\kappa}( r \leq r_1) -
\bar{\kappa}( r_2 < r \leq r_{max}) \nonumber \\ =  2 \int_{r_1} ^{r_2} \frac{
  \left < \gamma_T \right > }{r}d r + \frac{2}{1-r_2^2/r_{max}^2}
\int_{r_2}^{r_{max}} \frac{ \left <\gamma_T \right >}{r} d r,
\label{eqn_aperture_densitometry}
\end{eqnarray}
\noindent 
where $\langle \gamma_T \rangle$ is the azimuthally averaged
tangential shear, $r_1$ is the aperture radius, and $r_2$ and
$r_{max}$ are the inner- and the outer-radii of the annulus.
$\zeta_c(r_1,r_2,r_{max})$
provides a density contrast of the region inside $r<r_1$ with respect
to the control annulus $(r_2,r_{max})$. We choose $r_2=600\arcsec (\mytilde1.9~\mbox{Mpc})$ and
$r_{max}=800\arcsec (\mytilde2.5~\mbox{Mpc})$ for the control annulus and estimate the density
within this region by projecting our NFW fitting results (Table 2).
 The input to the equation of the densitometry is a shear, not a reduced shear. Therefore, we determine the aperture mass iteratively by updating $\kappa$.
The result is displayed in Figure~\ref{fig_aperture_mass}. Also, displayed in Figure~\ref{fig_aperture_mass} is the aperture mass estimated with the NFW fitting results above. In order to obtain this estimation, we first projected each NFW profile along the line-of-sight direction and sum the two cluster results. The projected masses from aperture mass densitometry and NFW fitting are in excellent agreement, which indicates that despite the apparent violent merger, parametrizing the halos with NFW profiles is still a good approximation at least  in \ciza.
 
Given that \ciza ~is comprised of two halos of approximately equal mass, we can estimate the total mass of the system in the following way.
We adopt the  mean position of the two halos as the cluster center. Namely, the coordinate
of this location is $(\alpha,\delta) \simeq(22^{h}42^{m}47^{s}.3,~53\degr01\arcmin 56\arcsec)$.
We set up a three-dimensional box (3D) with a side of 6 Mpc containing the two halos of \ciza. A density is assigned to
each volume element based on the parameters in Table 2. Then, it is straightforward to compute the total spherical
mass as a function of radius. We estimate that $M_{200c}=(2.51\pm0.53) \times 10^{15}~M_{\sun}$ is reached at $r_{200c}=2.63$~Mpc.
This mass is comparable to the value  $M_{200c}=(2.76\pm0.51) \times 10^{15}~M_{\sun}$    of ``El Gordo" similarly comprised of
two $\mytilde10^{15}M_{\sun}$ halos.
We remind readers that the total mass becomes greater than a mere sum of the two halos' masses because $r_{200c}$ should increase.
Although this total mass estimation assumes that the two subclusters are at the same distance from us, the result does not
vary significantly with viewing angles (see Jee et al. 2014b for the case of El Gordo) as long as the angle between the merger axis
and the plane of the sky is less than $\mytilde70\degr$. The polarization of the radio relic indicates that the angle is less than $\mytilde30\degr$ (van Weeren et al. 2010), which
is consistent with the result from our detailed dynamical analysis (Dawson et al. in prep.). 

\begin{deluxetable}{ccccc}
\tablewidth{0pt}
\tablecaption{Mass Estimates of \ciza~ based on 1D fit. \label{tab_mass1}}
\tablehead{  \colhead{Subclusters} & \colhead{$\sigma_v$}  & \colhead{$c_{200c}$}  & \colhead{$R_{200c}$} & \colhead{$M_{200c}$} \\
 & \colhead{(\kms)}  & \colhead{}  & \colhead{($\hubblem$ Mpc)} & \colhead{(\solarm)}  }
\startdata
North                                     & $1046\pm65$  &  $3.19\pm0.06$  &  $2.0\pm0.1$  & $11.1\pm2.2$ \\
South                                     & $1182\pm58$  &  $3.10\pm0.04$  &  $2.3\pm0.1$ &  $15.7\pm2.5$ 
\enddata
\tablecomments{The $\sigma_v$ value is derived from the SIS fit whereas the rest are from the NFW fit}
\end{deluxetable}

\begin{deluxetable}{ccccc}
\tablewidth{0pt}
\tablecaption{Mass Estimates of \ciza~ based on simultaneous 2D fit. \label{tab_mass2}}
\tablehead{  \colhead{Subclusters} & \colhead{$\sigma_v$}  & \colhead{$c_{200c}$}  & \colhead{$R_{200c}$} & \colhead{$M_{200c}$} \\
 & \colhead{(\kms)}  & \colhead{}  & \colhead{($\hubblem$ Mpc)} & \colhead{(\solarm)}  }
\startdata
North                                     & $967_{-128}^{+113}$  &  $3.20_{-0.08}^{+0.09}$  &  $2.0_{-0.2}^{+0.2}$  & $11.0_{-3.2}^{+3.7}$\\
South                                     & $1137_{-101}^{+93}$  &  $3.23_{-0.08}^{+0.08}$  &  $1.9_{-0.2}^{+0.1}$  & $9.8_{-2.5}^{+3.8}$
\enddata
\tablecomments{The $\sigma_v$ value is derived from the SIS fit whereas the rest are from the NFW fit}
\end{deluxetable}

\section{DISCUSSION} \label{section_discussion}

\subsection{Comparison with Dynamical and X-ray Studies}

To date, we have 217 spectroscopically confirmed members of \ciza. A comprehensive dynamical analysis will be presented in a separate publication (W. Dawson et al. in prep). Here we will present basic comparisons relevant to mass properties. 

Obviously, it is impossible to  distinguish the membership of galaxies individually between the two halos. Thus, we adopt a common operational definition based on the (projected) proximity to each clump. Within the $r=200\arcsec$ circle ($\mytilde 625$ kpc), we find that the number of galaxies belonging to the northern and southern components are 69 and 62, respectively.
The redshift of each component is determined with bi-weight estimator, giving $<z>=0.18794\pm0.00053$ and $0.18821\pm0.00052$ for the northern and southern clumps. The line-of-sight velocity difference between the two components is $\Delta z=0.00027$ ($\mytilde69\mbox{km}~\mbox{s}^{-1}$), which indicates that the merger is happening nearly in the plane of the sky.  

The directly measured velocity dispersions of the northern and southern clumps are $\sigma_v=1163_{-86}^{+103}~\mbox{km}~\mbox{s}^{-1}$ and $1084_{-74}^{+102}~\mbox{km}~\mbox{s}^{-1}$, respectively. These estimates can be compared with our lensing results by assuming that the cluster is composed of two singular isothermal spheres. Our simultaneous two-dimensional fitting (Table 2) gives  
$\sigma_v=967_{-128}^{+113}~\mbox{km}~\mbox{s}^{-1}$ and $1137_{-101}^{+93}~\mbox{km}~\mbox{s}^{-1}$ for the northern and southern components, respectively. The predicted  velocity dispersion from our lensing analysis for both subclusters are
consistent with the direct measurement.

One may choose to convert the directly measured velocity dispersion to a mass and compare the result to weak-lensing mass. Using the $M_{200c}-\sigma_{DM}$ relation of Evrard et al. (2008), we convert  $\sigma_v=1163_{-86}^{+103}~\mbox{km}~\mbox{s}^{-1}$ and $1084_{-74}^{+102}~\mbox{km}~\mbox{s}^{-1}$
to $M_{200c}=1.61_{-0.33}^{+0.46}\times10^{15}M_{\sun}$ and $1.30_{-0.25}^{+0.40}\times10^{15}M_{\sun}$, respectively, where the errors on the mass include only the statistical noise (not the scatter of the relation). These values are consistent with our
lensing results (Table 2).

The first gas temperature measurement of \ciza~ is presented in Ogrean et al. (2013). The result is based on deep XMM-Newton data. The global
temperature $7.7\pm0.3$ keV indicates that the cluster is massive.  Assuming that the
cluster follows an isothermal beta model with $\beta_X=0.7$ and $r_c=100$~kpc, we obtain $M_{200c}\sim1.3\times10^{15} M_{\sun}$.
This value is a factor of two smaller than our weak-lensing result, which is not surprising because the entire system \ciza~ cannot
be modeled with a single isothermal beta halo.

\subsection{Implication for Merging Scenario}

The giant radio relic of \ciza~ provides direct evidence for the on-going merger, which is also supported by the gas and galaxy distribution.
The numerical simulation of van Weeren et al. (2011) successfully reproduces the observed features of the radio relics in \ciza. Here
we discuss the implication of the current weak-lensing results on the initial condition by noting the following two apparent differences between our findings and the assumptions made for the simulation of van Weeren et al. (2011).

First, the most striking difference is the total mass of the system. Our weak-lensing analysis finds that the total mass of \ciza~ should exceed $M_{200c}\mytilde2\times10^{15} M_{\sun}$ (\textsection\ref{section_mass}).  On the other hand, in van Weeren et al. (2011)  the total mass was assumed to be $5.5\times10^{14} M_{\sun}$, which is a factor of 4 lower.

Second, the mass ratio is different. van Weeren et al. (2011) found that the mass ratio of the northern cluster to the southern cluster
should be close to 2:1 in order to reproduce the observed asymmetry of the two relics.  However, this 2:1 mass ratio is not favored by our weak-lensing analysis, which shows that the best-fit masses of the two sub-clusters are in fact of nearly equal mass.

The first difference (i.e., total mass) comes from the fact that van Weeren et al. (2011) assume that the total mass of \ciza~ is $5.5 \times 10^{14} M_{\sun}$, which is based on the luminosity-mass ($L_X-M$) relation of Pratt et al. (1998). 
One of the most important  merger parameters that will be affected by updating the mass would be the impact velocity between the two halos.
Using the simplified assumption made in Sarazin (2002), we can estimate the initial separation $d_0$ using Kepler's 3rd law
\begin{equation}
d_0 \sim 4.5 \left ( \frac{M_1 + M_2} {10^{15} M_{\sun}  } \right ) ^{1/3} \left ( \frac {t_{impact}} {10^{10} \mbox{yr} } \right ) ^{2/3} \mbox{Mpc}, 
\end{equation} 
\noindent
where $t_{impact}$ is the time elapsed at the impact since the unknown epoch $t_0$ when the two halos detached from the Hubble flow and started to 
accelerate toward each other. It is important to remember that the impact velocity that we discuss below is not sensitive to the exact value of $d_0$ ($t_0$).

It is impractical to set up a numerical simulation to begin at $t=t_0$ and run it until $t=t_{impact}$. Therefore, most $N$-body simulations choose a reasonable initial
distance $d$ as a starting point. The relative velocity between the two halos when they are separated by $d$ is given by (Sarazin 2002)
\begin{eqnarray}
v\sim 2930 \left (    \frac{M_1 + M_2} { 10^{15} M_{\sun} } \right )^{1/2} 
  \left ( \frac {1-d/d_0}{1-{ (b/d_0)^2} } \right )^{1/2} \times \nonumber \\  
  \left ( \frac{d}{1~\mbox{Mpc} }  \right )^{-1/2} \mbox{km}~\mbox{s}^{-1},
\end{eqnarray}
\noindent
where $b$ is the impact parameter.
When we assume $t_{impact} = t_{age}$ ($\mytilde 11$ Gyrs), $b=0$, and $M_1+M_2=5.5\times10^{14}$, the relative velocity of the two halos is 
$\mytilde 1,000~\mbox{km}~\mbox{s}^{-1}$ at $d=2$ Mpc. In fact, this relative velocity is used as an initial condition in the simulation of van Weeren et al. (2011).
Instead, if we substitute our weak-lensing masses into $M_1+M_2$,  the relative velocity increases to $\mytilde2,500~\mbox{km}~\mbox{s}^{-1}$ at the same separation $d=2$ Mpc. This difference is non-negligible and will greatly amplify at the core pass-through. The estimation of the exact
relative speed at the core impact requires $N$-body simulations, and we will defer the estimation to our future studies. 

The second difference may appear puzzling because the simulation clearly shows that a mass ratio of 2:1 (the northern clump should be more massive than
the southern clump by a factor of two) is required to reproduce the observed asymmetry in the two radio relics. Nevertheless, given
the size of the mass uncertainties in the current study, our result does not rule out the case of the 2:1 mass ratio. As shown in Figure~\ref{fig_c_and_m}, the tension is only slightly greater than  1~$\sigma$. In addition, even in the case that the global mass ratio is 1:1, the asymmetry in the relics of \ciza~can be accommodated in the scenario, wherein the two subcluster's concentration values are different. Since a halo with a higher concentration  packs more mass within the same volume (at a small radius), it may be possible to cause the observed disparity in radio relics if the northern cluster has a denser core.

\subsection{Mass-Galaxy Offset} \label{section_mass_galaxy_offset}

\begin{figure*}
\includegraphics[width=9cm]{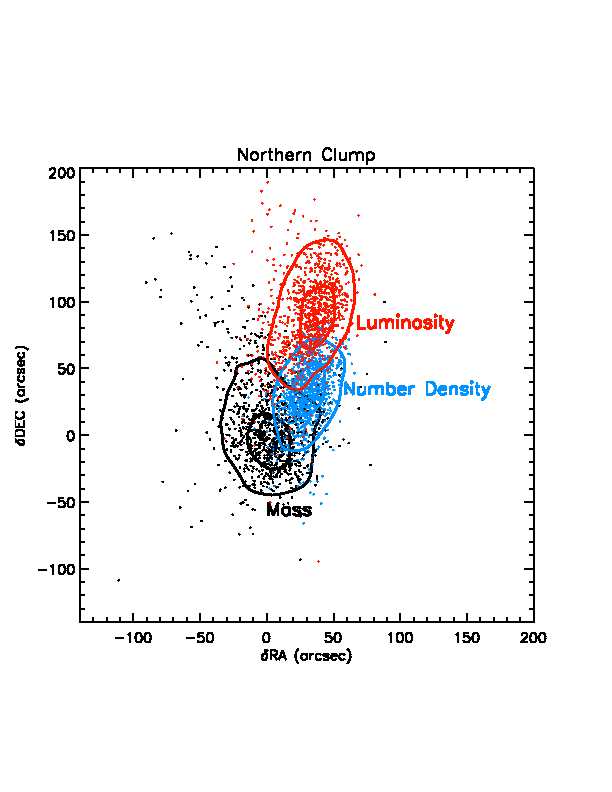}
\includegraphics[width=9cm]{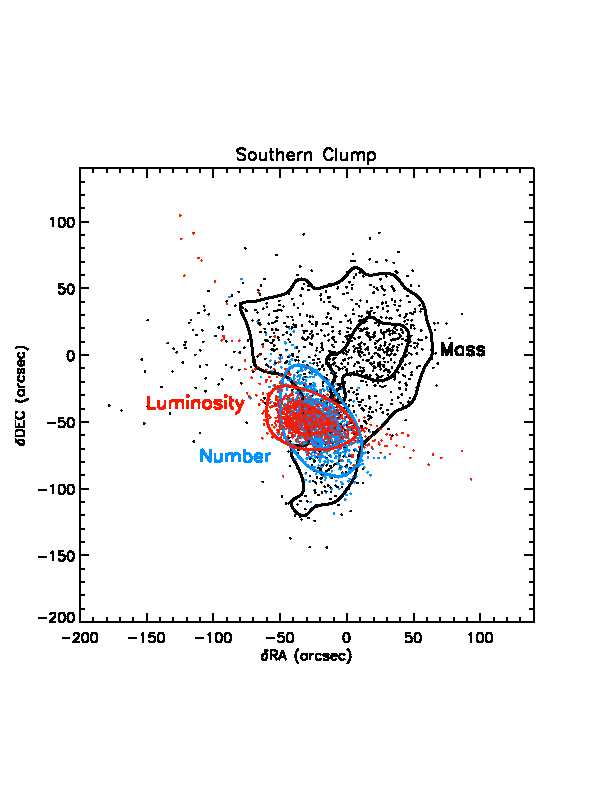}
\vspace{-2cm}
\caption{Centroid uncertainty estimation using bootstrap resampling. The coordinate (0,0) represents the approximate median 
of the mass centroid distribution. Each dot represents a measurement from a single realization (out of the 1000 boostrapping runs).
 The inner and outer contours corresponds to 1 and 2 $\sigma$ levels, respectively. The centroid distribution for each mass clump is similar to the shape in mass distribution. Accordingly, the northern mass clump, which is more peaked than the southern clump, has a smaller centroid uncertainty.
On the other hand, the size of the galaxy centroid distribution is similar between the northern and southern clusters.
Our bootstrapping analysis shows that both northern and southern
mass peaks are offset from the corresponding galaxy peaks at least at the $\mytilde2\sigma$ level;
the $1-\sigma$ contours of the mass and galaxy centroids do not overlap.}
\label{fig_centroid}
\end{figure*}

\begin{figure*}
\includegraphics[width=9cm]{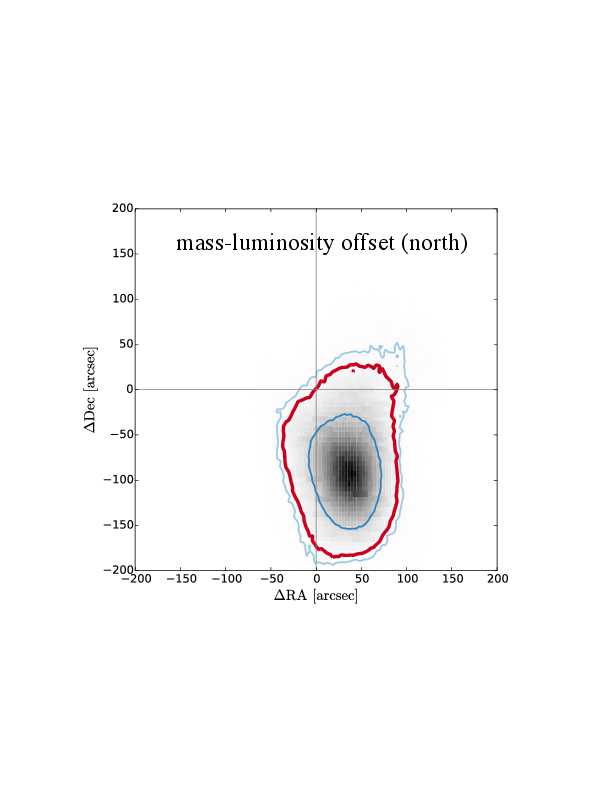}
\includegraphics[width=9cm]{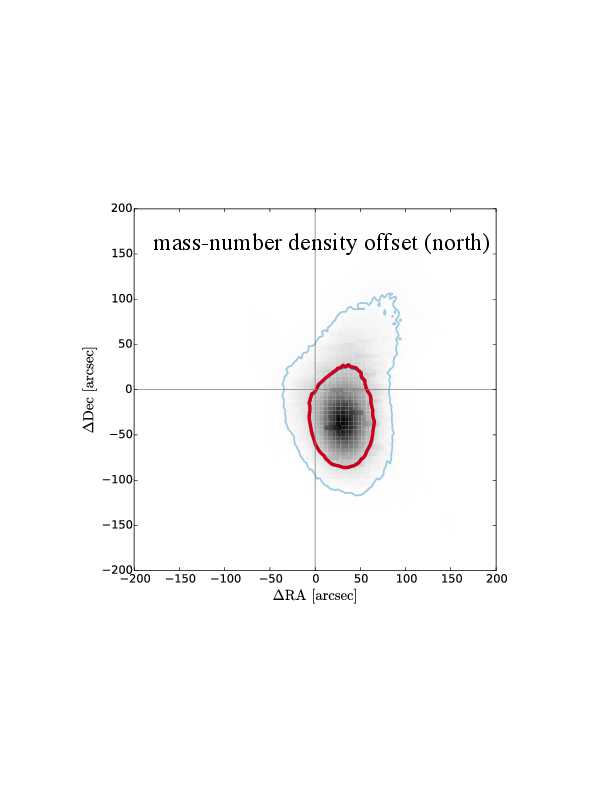}
\includegraphics[width=9cm]{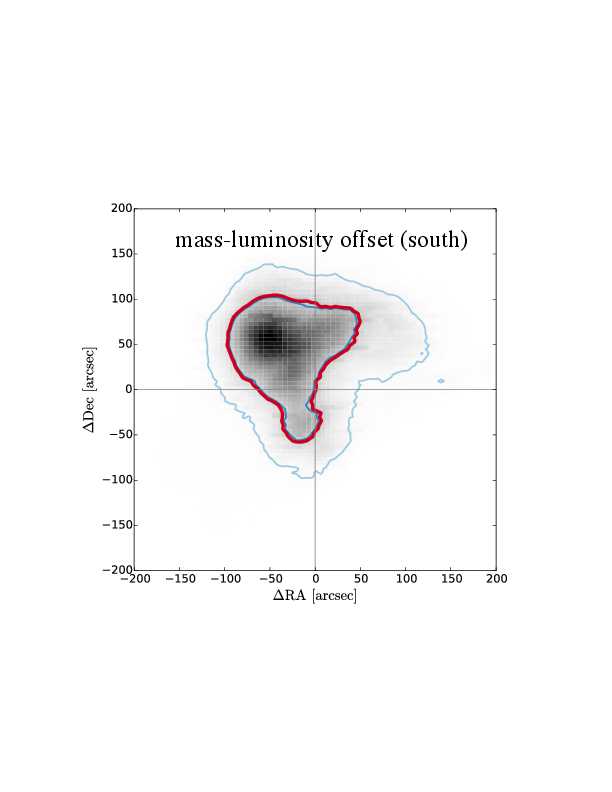}
\includegraphics[width=9cm]{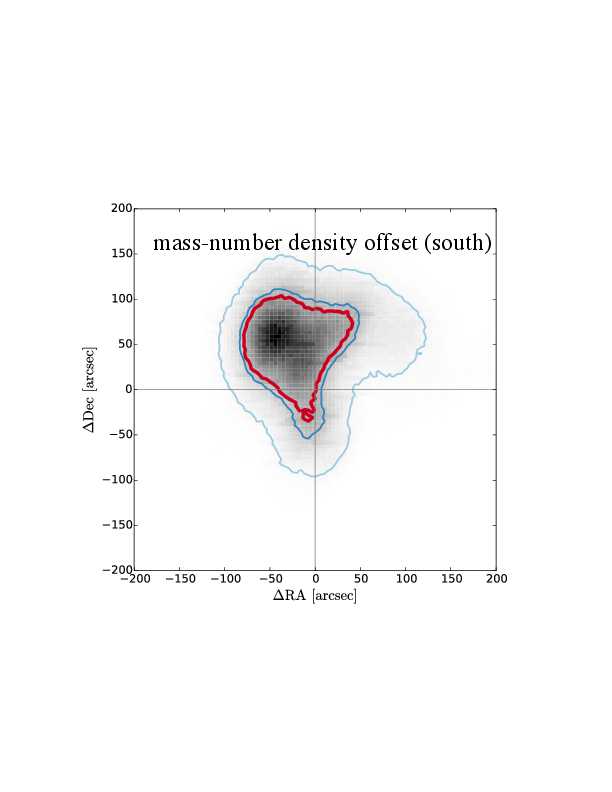}

\vspace{-2cm}
\caption{Significance of mass-galaxy offset. 
We examine the spatial distribution of the mass-galaxy offsets extracted from the bootstrap resampling.  The blue contours show the 68\% and 95\% confidence limits.  The origin in each plot corresponds to the null hypothesis of no offset, and the red contour shows the confidence limit that just touches this point in parameter space. The corresponding p-values are listed in Table 3.}
\label{fig_centroid_sig_test}
\end{figure*}

\begin{deluxetable}{ccc}
\tablewidth{0pt}
\tablecaption{Significance of the mass-galaxy offsets in \ciza. \label{tab_mg_offset}}
\tablehead{  \colhead{Subclusters} & \colhead{Mass-Luminosity}  & \colhead{Mass-Number} }
\startdata
North                                     & 0.082  &  0.310  \\
South                                     & 0.288  &  0.431
\enddata
\tablecomments{We quote $p$-values for the null hypothesis that the mass and galaxy centroids coincide.}
\end{deluxetable}

The comparison of the mass and galaxy centroids in \ciza ~shows that on average they are offset by $\mytilde1\arcmin$. 
The offset may be regarded as interesting in the context of a SIDM theory with a non-negligible collisional cross-section, which predicts that in general mass and galaxy distributions will
differ after collision in merging clusters. In \ciza ~ the northern galaxy centroid appears to lead the corresponding mass centroid whereas the mass-galaxy offset is rather lateral
for the southern clump. Also, the axis defined by the line connecting the two mass centroids is rotated by $\sim10\degr$ with respect to the merger axis defined by the
cluster mass, X-ray gas, and radio relics.

In order to investigate the significance of the mass centroid, we re-sample source galaxies while allowing duplication. Similarly, we bootstrap the cluster galaxies as well to estimate the uncertainty of their centroid in both luminosity and number density.
The resulting centroid distributions are displayed in Figure~\ref{fig_centroid}. We note that the luminosity centroid uncertainty shown here should be considered
an upper limit because we assign equal probabilities of elimination/selection to all galaxies regardless of their luminosity; a tighter centroid distribution would be
obtained if brighter galaxies were made harder to drop in the re-sampled catalog.

Before we discuss the mass-galaxy offset, it is instructive to examine individual contours in detail.
First, we note that the mass centroid contours closely resemble the shapes of the mass peaks (Figure~\ref{fig_massovergal}). 
Second, the northern mass peak, which is more peaked, has a tighter centroid distribution.
Third, the galaxy centroid distributions also resemble the corresponding smoothed galaxy (both luminosity and number density) distribution.
Fourth, the size of the galaxy centroid uncertainties are similar between the northern and southern clumps. 

The significance of the mass-galaxy offset is not easily deduced from Figure~\ref{fig_centroid}. One question is whether number density or luminosity is a better tracer of the galaxy population.  Luminosity is a better tracer of the mass in stars in principle, but its location can be more highly disturbed by interloping foreground galaxies. We therefore tabulate both types of offset in both north and south subclusters, yielding four potential offsets.  In Figure~\ref{fig_centroid_sig_test},  we plot the spatial distribution of these four types of offsets extracted from the bootstrap resampling. The blue contours show the 68\% and 95\% confidence limits.  The origin in each plot corresponds to the null hypothesis of no offset, and the red contour shows the confidence limit that just touches this point in parameter space.  The corresponding p-values are listed in Table 3. The lowest p-value is seen in the mass-luminosity offset in the north, meaning that the probability of detecting the observed offset by a random chance is $\mytilde8$\%.

Can we use the current mass-galaxy (luminosity) offset for the argument of SIDM? We believe that it is still premature to conclude that our observation supports SIDM because of the following caveats.
First, we do not know the ICM mass exactly. As seen in Figure~\ref{fig_ciza_first}, the X-ray emission is
more centrally concentrated than the cluster galaxies, which may imply that a significant fraction of ICM
may be present between the two halos of \ciza. Hence, one can suspect that our mass centroids
are somewhat pulled inward (toward the cluster center) if the fraction of the mass locked up in the ICM is significant.
Dawson (2013) showed that for the Musket Ball Cluster (Dawson et al. 2012), this systematic offset
can be of the same order of magnitude as the weak-lensing and galaxy centroid uncertainties.
Second, we cannot yet exclude the possibility that any line-of-sight interlopers are present. This issue
may be resolved in the future when we reach more completeness in our spectroscopic survey of the field.
Currently, our existing spectroscopic catalog does not hint at the possibility.
Third, our weak-lensing source catalog may be ``asymmetrically" contaminated by the cluster galaxies.
With our dependence on the color-magnitude relation in source selection, we expect that our source catalog
contains some blue cluster galaxies inevitably. Typically, we assume that the contamination
is azimuthally symmetric in isolated clusters. However, in clusters with elongated morphology like \ciza, the assumption
may turn out to be incorrect, especially, when the merger triggers star-formation activity preferentially along the merger axis
in the outskirts (Stroe et al. 2014b). 
Fourth, interpretation of galaxy ellipticity is ambiguous in the strong lensing regime. That is,
we do not know whether a galaxy ellipticity $e$ becomes $g$ or $1/g^*$ a priori. The two
cases can be reliably distinguished only when strong lensing data with known redshifts are used.
Because \ciza~ is a merging cluster, a probable shape of the critical curve is a long ellipse elongated in the north-south direction
similar to the one modeled by Zitrin et al. (2013) for ``El Gordo".
Therefore, it is worth investigating in the future how weak-lensing mass centroids are affected in a similar mass configuration.

\section{CONCLUSIONS} \label{section_conclusion}

We have presented detailed weak gravitational lensing analysis of \ciza, which is a rare target boasting a textbook example of a giant radio relic.
Our mass reconstruction shows that the cluster consists of two halos of nearly equal mass, consistent with the cluster galaxy distribution. In order to quantify the mass while minimizing the contamination bias, we perform MCMC by simultaneously fitting the two halos with NFW profiles.
Although we adopt the luminosity centers as our initial values, the mass centroids are not fixed, but allowed to move during our MCMC run.
We obtain  $M_{200c}=11.0_{-3.2}^{+3.7}\times10^{14}~M_{\sun}$ and $9.8_{-2.5}^{+3.8}\times10^{14}~M_{\sun}$ for
the northern and southern halos, respectively. The total mass of the system is $M_{200c}=(2.51\pm0.53)\times10^{15}~M_{\sun}$ at $r_{200c}=2.63$~Mpc, which makes \ciza ~among the most massive clusters detected so far. 
Because of the low redshift of the cluster, this mass alone does not cause any tension with the $\Lambda$CDM paradigm.
This large cluster mass suggests that the previous $N$-body simulation assuming the total mass $\mytilde5.5\times10^{14}M_{\sun}$ 
should be revised. 

We detect somewhat large offsets ($\mytilde1\arcmin$) between galaxy and mass centroids. To quantify the significance, we run bootstrapping for both mass and galaxy distributions and find that the offsets are significant at least at the 2 $\sigma$ level for the northern clump. Although the direction of the offset is
in favor of SIDM at face value, we identify some caveats that may lead to the current observation. They are 1) the gas mass that can pull
the total mass (galaxy+gas+dark matter) centroids toward the cluster center with respect to the galaxies, 2) a possible line-of-sight structure that is uncorrelated with the cluster, but may bias our mass centroids,  3) the cluster galaxy contamination in our source catalog, whose contamination rate may differ azimuthally and thus shift the two weak-lensing mass centroids, and 4) the ambiguity of shear interpretation between strong
and weak lensing regimes.
We believe that all these four issues can be addressed in the future as more concerted efforts are made to enhance our understanding of the system.

We thank the Merging Cluster Collaboration for useful discussions and comments.
MJJ, DW, and WD acknowledge support from HST-GO-13343.01-A.
Part of this work performed under the auspices of the U.S. DOE by LLNL under Contract DE-AC52-07NA27344.
AS acknowledges financial support from NWO. MB acknowledges funding from the Deutsche Forschungsgemeinschaft under SFB 676. DS acknowledges financial support from the Netherlands Organisation for Scientific research (NWO) through a Veni fellowship, from FCT through a FCT Investigator Starting Grant and Start-up Grant (IF/01154/2012/CP0189/CT0010) and from FCT grant PEst-OE/FIS/UI2751/2014.
RW is supported by NASA through the Einstein Postdoctoral grant number PF2-130104 awarded by the Chandra X-ray Center, which is operated by the Smithsonian Astrophysical Observatory for NASA under contract NAS8-03060.


\begin{thebibliography}{}
\bibitem[Abazajian et al.(2009)]{2009ApJS..182..543A} Abazajian, K.~N., 
Adelman-McCarthy, J.~K., Ag{\"u}eros, M.~A., et al.\ 2009, \apjs, 182, 543 
\bibitem[Akamatsu 
\& Kawahara(2013)]{2013PASJ...65...16A} Akamatsu, H., \& Kawahara, H.\ 2013, \pasj, 65, 16 
\bibitem[Bertin(2006)]{2006ASPC..351..112B} Bertin, E.\ 2006, Astronomical 
Data Analysis Software and Systems XV, 351, 112 
\bibitem[Bertin et al.(2002)]{2002ASPC..281..228B} Bertin, E., Mellier, Y., 
Radovich, M., et al.\ 2002, Astronomical Data Analysis Software and Systems 
XI, 281, 228 
\bibitem[Clowe et al.(2000)]{2000ApJ...539..540C} Clowe, D., Luppino,
G.~A., Kaiser, N., \& Gioia, I.~M.\ 2000, \apj, 539, 540
\bibitem[Clowe et al.(2006)]{2006ApJ...648L.109C} Clowe, D., Brada{\v c},
M., Gonzalez, A.~H., et al.\ 2006, \apjl, 648, L109
\bibitem[Dawson et al.(2012)]{2012ApJ...747L..42D} Dawson, W.~A., Wittman,
D., Jee, M.~J., et al.\ 2012, \apjl, 747, L42
\bibitem[Dawson(2013)]{2013PhDT.......211D} Dawson, W.~A.\ 2013, 
Ph.D.~Thesis
\bibitem[Duffy et al.(2008)]{2008MNRAS.390L..64D} Duffy, A.~R., Schaye, J.,
Kay, S.~T., \& Dalla Vecchia, C.\ 2008, \mnras, 390, L64
\bibitem[Ensslin et 
al.(1998)]{1998A&A...332..395E} Ensslin, T.~A., Biermann, P.~L., Klein, U., \& Kohle, S.\ 1998, \aap, 332, 395 
\bibitem[Evrard et al.(2008)]{2008ApJ...672..122E} Evrard, A.~E., Bialek, 
J., Busha, M., et al.\ 2008, \apj, 672, 122 
\bibitem[Fahlman et al.(1994)]{fahlman94} Fahlman,G.,Kaiser,N.,Squires,G.\& Woods,D. 1994,\apj, 437, 56
\bibitem[Ilbert et al.(2009)]{2009ApJ...690.1236I} Ilbert, O., Capak, P., 
Salvato, M., et al.\ 2009, \apj, 690, 1236 
\bibitem[Jee et al.(2007)]{2007PASP..119.1403J} Jee, M.~J., Blakeslee,
J.~P., Sirianni, M., Martel, A.~R., White, R.~L., \& Ford, H.~C.\ 2007a, \pasp, 119, 1403
\bibitem[Jee et al.(2007)]{2007ApJ...661..728J} Jee, M.~J., et al.\ 2007b,
\apj, 661, 728
\bibitem[Jee
\& Tyson(2011)]{2011PASP..123..596J} Jee, M.~J., \& Tyson, J.~A.\ 2011, \pasp, 123, 596
\bibitem[Jee et al.(2012)]{2012ApJ...747...96J} Jee, M.~J., Mahdavi, A.,
Hoekstra, H., et al.\ 2012, \apj, 747, 96
\bibitem[Jee et al.(2013)]{2013ApJ...765...74J} Jee, M.~J., Tyson, J.~A., 
Schneider, M.~D., et al.\ 2013, \apj, 765, 74 
\bibitem[Jee et al.(2014)]{2014ApJ...783...78J} Jee, M.~J., Hoekstra, H., 
Mahdavi, A., \& Babul, A.\ 2014a, \apj, 783, 78 
\bibitem[Jee et al.(2014)]{2014ApJ...785...20J} Jee, M.~J., Hughes, J.~P., 
Menanteau, F., et al.\ 2014b, \apj, 785, 20 
\bibitem[Kaiser
\& Squires(1993)]{1993ApJ...404..441K} Kaiser, N., \& Squires, G.\ 1993, \apj, 404, 441
\bibitem[Katgert et 
al.(1973)]{1973A&A....23..171K} Katgert, P., Katgert-Merkelijn, J.~K., Le Poole, R.~S., \& van der Laan, H.\ 1973, \aap, 23, 171 
\bibitem[Kocevski et al.(2007)]{2007ApJ...662..224K} Kocevski, D.~D., 
Ebeling, H., Mullis, C.~R., \& Tully, R.~B.\ 2007, \apj, 662, 224 
\bibitem[Mahdavi et al.(2007)]{2007ApJ...668..806M} Mahdavi, A., Hoekstra,
H., Babul, A., Balam, D.~D., \& Capak, P.~L.\ 2007, \apj, 668, 806
\bibitem[Menanteau et al.(2012)]{2012ApJ...748....7M} Menanteau, F.,
Hughes, J.~P., Sif{\'o}n, C., et al.\ 2012, \apj, 748, 7
\bibitem[Merten et al.(2011)]{2011MNRAS.417..333M} Merten, J., Coe, D.,
Dupke, R., et al.\ 2011, \mnras, 417, 333
\bibitem[Navarro et al.(1997)]{1997ApJ...490..493N} Navarro, J.~F., Frenk, 
C.~S., \& White, S.~D.~M.\ 1997, \apj, 490, 493 
\bibitem[Ogrean et al.(2013)]{2013MNRAS.429.2617O} Ogrean, G.~A., 
Br{\"u}ggen, M., R{\"o}ttgering, H., et al.\ 2013, \mnras, 429, 2617 
\bibitem[Ogrean et al.(2014)]{2014MNRAS.440.3416O} Ogrean, G.~A., 
Br{\"u}ggen, M., van Weeren, R., et al.\ 2014, \mnras, 440, 3416 
\bibitem[Ouchi et al.(2004)]{2004ApJ...611..660O} Ouchi, M., Shimasaku, K., 
Okamura, S., et al.\ 2004, \apj, 611, 660 
\bibitem[Pratt et al.(2009)]{2009A&A...498..361P} Pratt, G.~W., Croston, J.~H., Arnaud, M., B\"{o}hringer, H.\ 2009, \aap, 498, 361 
\bibitem[Rocha et al.(2013)]{2013MNRAS.430...81R} Rocha, M., Peter, 
A.~H.~G., Bullock, J.~S., et al.\ 2013, \mnras, 430, 81 
\bibitem[Sarazin(2002)]{2002ASSL..272....1S} Sarazin, C.~L.\ 2002, Merging 
Processes in Galaxy Clusters, 272, 1 
\bibitem[Schlafly 
\& Finkbeiner(2011)]{2011ApJ...737..103S} Schlafly, E.~F., \& Finkbeiner, D.~P.\ 2011, \apj, 737, 103 
\bibitem[Seitz
\& Schneider(1997)]{1997A&A...318..687S} Seitz, C., \& Schneider, P.\ 1997, \aap, 318, 687
\bibitem[Skrutskie et al.(2006)]{2006AJ....131.1163S} Skrutskie, M.~F., 
Cutri, R.~M., Stiening, R., et al.\ 2006, \aj, 131, 1163 
\bibitem[Stroe et al.(2013)]{2013A&A...555A.110S} Stroe, A., van Weeren, R.~J., Intema, H.~T., et al.\ 2013, \aap, 555, A110 
\bibitem[Stroe et al.(2014)]{2014MNRAS.441L..41S} Stroe, A., Rumsey, C., 
Harwood, J.~J., et al.\ 2014a, \mnras, 441, L41 
\bibitem[Stroe et al.(2014)]{2014MNRAS.438.1377S} Stroe, A., Sobral, D., 
R{\"o}ttgering, H.~J.~A., \& van Weeren, R.~J.\ 2014b, \mnras, 438, 1377 
\bibitem[Swarup et al.(1991)]{1991CuSc...60...95S} Swarup, G., 
Ananthakrishnan, S., Kapahi, V.~K., et al.\ 1991, Current Science, Vol.~60, 
NO.2/JAN25, P.~95, 1991, 60, 95 

\bibitem[van Weeren et al.(2010)]{2010Sci...330..347V} van Weeren, R.~J., 
R{\"o}ttgering, H.~J.~A., Br{\"u}ggen, M., 
\& Hoeft, M.\ 2010, Science, 330, 347 
\bibitem[van Weeren et al.(2011)]{2011MNRAS.418..230V} van Weeren, R.~J., 
Br{\"u}ggen, M., R{\"o}ttgering, H.~J.~A., 
\& Hoeft, M.\ 2011, \mnras, 418, 230 
\bibitem[Yagi et al.(2002)]{2002AJ....123...66Y} Yagi, M., Kashikawa, N., 
Sekiguchi, M., et al.\ 2002, \aj, 123, 66 
\bibitem[Zwart et al.(2008)]{2008MNRAS.391.1545Z} Zwart, J.~T.~L., Barker, 
R.~W., Biddulph, P., et al.\ 2008, \mnras, 391, 1545 
\end{thebibliography}
\end{document}